\documentclass[%
 reprint,
superscriptaddress,
 amsmath,amssymb,
 aps,
 prl,
]{revtex4-1}

\usepackage{booktabs}
\usepackage{enumitem}
\usepackage{graphicx}
\usepackage{hyperref}
\usepackage[utf8x]{inputenc}
\usepackage{siunitx}
\usepackage{subcaption}
\usepackage{supertabular}

\makeatletter
\renewcommand\@makecaption[2]{%
  \par
  \vskip\abovecaptionskip
  \begingroup
   \small\rmfamily
    \begingroup
     \samepage
     \flushing
     \let\footnote\@footnotemark@gobble
     \@make@capt@title{#1}{#2}\par
    \endgroup
  \endgroup
  \vskip\belowcaptionskip
}
\makeatother

\newcommand{\PX}{PandaX-4T~}

\newcommand{\Geantbased}{G\scriptsize{EANT}\normalsize{4-based}~}

\newcommand{\btm}[1]{\ensuremath{#1}$_{\rm b}$}

\newcommand{\GeV}{GeV/$c^2$\,}
\newcommand{\TeV}{TeV/$c^2$\,}

\NewDocumentCommand{\ncl}{ >{\SplitArgument{1}{-}}m }{\nclaux#1}
\NewDocumentCommand{\nclaux}{mm}{\ensuremath{\!{}^{#2}\mathrm{#1}}}
\NewDocumentCommand{\nclm}{ >{\SplitArgument{1}{-}}m }{\nclmaux#1}
\NewDocumentCommand{\nclmaux}{mm}{\ensuremath{\!{}^{#2\mathrm{m}}\mathrm{#1}}}

\makeatletter

\newcommand{\Rmnum}[1]{\expandafter\@slowromancap\romannumeral #1@}
\makeatother

\begin{document}
\title{Dark Matter Search Results from 1.54 Tonne$\cdot$Year Exposure of \PX}


\def\shKeyLab{School of Physics and Astronomy, Shanghai Jiao Tong University, Key Laboratory for Particle Astrophysics and Cosmology (MoE), Shanghai Key Laboratory for Particle Physics and Cosmology, Shanghai 200240, China}
\def\scKeyLab{Jinping Deep Underground Frontier Science and Dark Matter Key Laboratory of Sichuan Province}
\def\BUAA{School of Physics, Beihang University, Beijing 102206, China}
\def\BUAACenter{Peng Huanwu Collaborative Center for Research and Education, Beihang University, Beijing 100191, China}
\def\BUAALab{Beijing Key Laboratory of Advanced Nuclear Materials and Physics, Beihang University, Beijing, 102206, China}
\def\SCNT{Southern Center for Nuclear-Science Theory (SCNT), Institute of Modern Physics, Chinese Academy of Sciences, Huizhou 516000, China}
\def\USTClab{State Key Laboratory of Particle Detection and Electronics, University of Science and Technology of China, Hefei 230026, China}
\def\USTCdep{Department of Modern Physics, University of Science and Technology of China, Hefei 230026, China}
\def\BUAALab{International Research Center for Nuclei and Particles in the Cosmos \& Beijing Key Laboratory of Advanced Nuclear Materials and Physics, Beihang University, Beijing 100191, China}
\def\pku{School of Physics, Peking University, Beijing 100871, China}
\def\YaLongSD{Yalong River Hydropower Development Company, Ltd., 288 Shuanglin Road, Chengdu 610051, China}
\def\IAP{Shanghai Institute of Applied Physics, Chinese Academy of Sciences, 201800 Shanghai, China}
\def\CHEPpku{Center for High Energy Physics, Peking University, Beijing 100871, China}
\def\SDUdep{Research Center for Particle Science and Technology, Institute of Frontier and Interdisciplinary Science, Shandong University, Qingdao 266237, Shandong, China}
\def\SDUlab{Key Laboratory of Particle Physics and Particle Irradiation of Ministry of Education, Shandong University, Qingdao 266237, Shandong, China}
\def\UMD{Department of Physics, University of Maryland, College Park, Maryland 20742, USA}
\def\TDLee{New Cornerstone Science Laboratory, Tsung-Dao Lee Institute, Shanghai Jiao Tong University, Shanghai 201210, China}
\def\MESJTU{School of Mechanical Engineering, Shanghai Jiao Tong University, Shanghai 200240, China}
\def\SYU{School of Physics, Sun Yat-Sen University, Guangzhou 510275, China}
\def\SYUSFI{Sino-French Institute of Nuclear Engineering and Technology, Sun Yat-Sen University, Zhuhai, 519082, China}
\def\NKU{School of Physics, Nankai University, Tianjin 300071, China}
\def\YTU{Department of Physics, Yantai University, Yantai 264005, China}
\def\FDU{Key Laboratory of Nuclear Physics and Ion-beam Application (MOE), Institute of Modern Physics, Fudan University, Shanghai 200433, China}
\def\USST{School of Medical Instrument and Food Engineering, University of Shanghai for Science and Technology, Shanghai 200093, China}
\def\SJTUSC{Shanghai Jiao Tong University Sichuan Research Institute, Chengdu 610213, China}
\def\SPEIT{SJTU Paris Elite Institute of Technology, Shanghai Jiao Tong University, Shanghai, 200240, China}
\def\NNU{School of Physics and Technology, Nanjing Normal University, Nanjing 210023, China}
\def\SYSUzhuhai{School of Physics and Astronomy, Sun Yat-Sen University, Zhuhai 519082, China}
\def\CDUT{College of Nuclear Technology and Automation Engineering, Chengdu University of Technology, Chengdu 610059, China}

\affiliation{\TDLee}
\author{Zihao Bo}\affiliation{\shKeyLab}
\author{Wei Chen}\affiliation{\shKeyLab}
\author{Xun Chen}\affiliation{\TDLee}\affiliation{\shKeyLab}\affiliation{\SJTUSC}\affiliation{\scKeyLab}
\author{Yunhua Chen}\affiliation{\YaLongSD}\affiliation{\scKeyLab}
\author{Zhaokan Cheng}\affiliation{\SYUSFI}
\author{Xiangyi Cui}\affiliation{\TDLee}
\author{Yingjie Fan}\affiliation{\YTU}
\author{Deqing Fang}\affiliation{\FDU}
\author{Zhixing Gao}\affiliation{\shKeyLab}
\author{Lisheng Geng}\affiliation{\BUAA}\affiliation{\BUAACenter}\affiliation{\BUAALab}\affiliation{\SCNT}
% ...
\collaboration{PandaX Collaboration}
\noaffiliation

\date{\today}

\begin{abstract}
In this Letter, we report the dark matter search results from the commissioning run (Run0) and the first science run (Run1) of the \PX experiment.
The two datasets were processed with a unified procedure, with the Run1 data treated blindly.
The data processing is improved compared to previous work, unifying the low-level signal reconstruction in a wide energy range up to 120 keV.
With a total exposure of 1.54~tonne$\cdot$year, no significant excess of nuclear recoil events is found.
The lowest 90\% confidence level exclusion on the spin-independent cross section is $1.6 \times 10^{-47}~\mathrm{cm}^2$ at a dark matter mass of $40$~\GeV.
Our results represent the most stringent constraint for a dark matter mass above $100$~\GeV.
\end{abstract}

\maketitle

\paragraph{Introduction—} Overwhelming astrophysical and cosmological evidence points to the existence of dark matter (DM) in our Universe~\cite{Bertone:2016nfn}.
Direct detection of dark matter particles through their potential interaction with ordinary matter is pivotal to our understanding of the fundamental nature of DM. 
Among many experimental endeavors, liquid xenon time projection chambers (TPCs) have spearheaded the detection sensitivities~\cite{Meng2021_P4First, Aalbers2023_LZFirst, Aprile2023_nTFirst}, with their sensitivity steadily approaching the irreducible neutrino-nucleus scattering background~\cite{Billard2014_NeutrinoFloor, O’Hare2021_NeutrinoFog}.

\paragraph{Experimental setup—} The \PX experiment~\cite{Zhang2019_P4Projection} is located in the B2 hall of the China Jinping Underground Laboratory (CJPL-II)~\cite{Kang2010, Li2014}.
The xenon TPC is inside a dual-vessel stainless steel cryostat, which holds 5.6 tonnes of total liquid xenon. The cryostat is housed in a 10-meter diameter by 13-meter height stainless steel tank, which is filled with ultrapure water to provide radioactivity shielding.
The xenon cooling and purification system comprises three cooling heads and two independent recirculation loops, continuously removing the electronegative impurities via hot getters~\cite{Zhao:2020vxh}. 
The detector features a cylindrical, dual-phase xenon TPC delineated by 24 reflective polytetrafluoroethylene (PTFE) panels, with a 1185 mm separation across opposing panels at room temperature.
Its internal electrical fields are established through a cathode grid at the bottom, and a gate and anode mesh positioned immediately below and above the liquid level, with a vertical spacing of 1185~mm and 10~mm, respectively.
Energy depositions produce prompt scintillation ($S1$) and delayed electroluminescence ($S2$) photons, which are detected by a total of 368 Hamamatsu R11410-23 3-inch photomultiplier tubes (PMTs) positioned at the top and bottom of the TPC. Each PMT's waveform is read out by a digitizer channel running at a sampling rate of 250 MS/s, with a self-trigger threshold set at around 1/3 of a photoelectron~\cite{Yang:2021hnn}. The $S1$-$S2$ detection allows precise energy and position reconstruction of the events.
The ratio $S2/S1$ also provides excellent discrimination between electron recoil (ER) and nuclear recoil (NR) events.
More detailed information on \PX configurations and operation can be found in Ref.~\cite{Meng2021_P4First} and the associated references therein.

The \PX commissioning run contains 95 days of stable data taking operation from November 28, 2020, to April 16, 2021 (Run0), separated into five sets according to different detector conditions.
Multiple analyses of the Run0 data have been conducted to search for various DM model particles~\cite{Meng2021_P4First, Gu2022_P4NRAbsp, Zhang2022_P4ERAbsp, Ning2023_EFT}.
To reduce the tritium background identified in Run0, a dedicated offline xenon distillation~\cite{Cui2021_P4Distillation, Cui2024_RnRemove} was carried out after Run0.
\PX resumed operations in the summer of 2021, collecting over 164 days of scientific data from November 16, 2021, to May 15, 2022 (Run1). Afterwards, the operation was halted until the end of 2023 due to ongoing construction at CJPL-II. In this Letter, we present a combined DM analysis using data from both Run0 and Run1.

The high voltage applied on the gate and cathode is $-6$ kV and $-16$ kV throughout Run1 data taking.
Although most of the other detector configurations remained the same as in Run0, several experimental issues occurred during Run1. In total, an additional fifteen PMT channels malfunctioned: eight due to a short in a shared negative high-voltage supply~\cite{Elsied:2015ixa}, and the remaining seven were turned off due to high afterpulses and dark rates. 
The shorted eight channels are adjacent to each other on the top array, and are also in the vicinity of three dead channels that failed earlier in Run0. Events in this region suffer from non-negligible charge loss, therefore special measures were taken in the data selection. 
Additionally, liquid level control failed throughout Run1. As a result, each time the recirculation conditions changed, the induced liquid level variations led to a (correctable) difference in the amplification of $S2$ signals.
Accordingly, Run1 data is divided into six subsets.

\paragraph{Data analysis details—}
In this analysis, a unified analysis procedure is applied to the Run0 and Run1 data, with Run1 data completely blinded, including the following major improvements:

Unified signal reconstruction procedure: 
Signal reconstruction refers to the offline identification and reconstruction of $S1$ and $S2$ signals from raw data, and the building of physical ``events'' to become basic units of the analysis. 
We adopt the ``rolling gain'' method instead of using the PMT gains obtained from weekly calibration with light-emitting diodes \cite{Yan2024_Xe134DBD}.
As previously, after hit clustering, the $S1$ and $S2$ signals are identified based on the number of hits, the charge ratio between the top and bottom arrays, and the signal width.
To discriminate $S1$s from fragments of single electron ionization $S2$ signals, we impose a series of requirements on $S1$ when paired with $S2$, including criteria on the $S1$ pulse shape profile, light pattern topology, and noise levels in PMT waveforms immediately before and after.
The event building is now performed with a 2-ms fixed window length (maximum electron drift time is about 900~$\mu$s).
This allows more consistent event level parameters, e.g. waveform noise parameter, in comparison to the dynamic event window used in Ref.~\cite{Meng2021_P4First}, and further eases the selection efficiency evaluation.
The improved signal reconstruction procedure is consistently applied to data from the low energy threshold up to 120~keV in electron equivalent energy, to facilitate future analyses on solar $pp$ neutrinos, etc.

Blinded data selection: 
Data selection criteria are set ``blindly''.
We conduct dedicated optimizations on data selection cuts based solely on all low-energy ER and NR calibration data~\cite{Luo2024_P4SignalModel}.
We eliminate periods with high $S1$ rates and veto the ``afterglow'' time window after large signals (varying with the signal size), resulting in a reduction of approximately 12\% of the exposure.
At the signal level, the data quality cuts include those related to the $S1$ and $S2$ pulse shapes, as well as the charge distribution on PMTs.
Furthermore, the $S1$ top-bottom charge partition is required to be consistent with the vertical position of the event, and the width of $S2$ should correlate with the drift time according to diffusion effect.
For Run 1, these cuts are relaxed if $S2$ is reconstructed close to the malfunctioned PMTs [green dashed regions in later Fig.~\ref{fig:event-dist-pos}(a)].
At the event level, events with excessive noise in the waveform are further removed.
To identify genuine single-site (SS) events, instead of simply counting the number of $S2$'s, multiple waveform cuts are developed and applied. 
For the DM search, we set the range of corrected $S1$ to be [2, 135] PE (with at least two PMTs in coincidence) and that for $S2$ to be [120, 20 000] PE, respectively.
This corresponds to an average ER (NR) energy window from 1 (4) to 25 (94) keV for Run0 and 1 (3) to 26 (103) keV for Run1. 
Candidate events are also required to be above the $99.5\%$ NR quantile to further suppress surface background and multisite (MS) events depositing their energy partially below the cathode plane.
The efficiencies and associated uncertainties in data selection are established with calibration data, and cross checked with our signal response model and a waveform simulation, which includes the effect of SS classification~\cite{Luo2024_P4SignalModel, Li2024_WfSim}.


Improved spatial and temporal signal corrections: 
The event position reconstruction mostly follows the methods developed in Refs.~\cite{Zhang2021_PosRecon, Meng2021_P4First}.
An extra azimuth-angle-dependent scaling correction is applied based on the uniformly distributed \,\nclm{Kr-83} ER events~\cite{Luo2024_P4SignalModel}. 
Within the fiducial volume (FV), the uniformity of \,\nclm{Kr-83} events in four quadrants is within $\pm8\%$, reflecting the impact of malfunctioned PMTs on the position reconstruction.
A series of $S1$ and $S2$ charge corrections vs event position are performed, using the {\it in situ}
(\,\ncl{Rn-222} 5.6~MeV $\alpha$s) and injected (\,\nclm{Kr-83} $41.5$ keV internal conversion electrons) radioactive sources, as described in Ref.~\cite{Luo2024_P4SignalModel}. 
To avoid systematic effects due to dead channels and saturation of the top PMTs, bottom-only $S2$ (i.e., \btm{S2}) is used for energy reconstruction.
The electron-equivalent energy $E_\mathrm{ee}$ of a SS event in \PX is reconstructed as
\begin{equation}
    E_\mathrm{ee} = W_q \left(\frac{Q^c_{S1}}{g_1}+\frac{Q^c_{S2_\mathrm{b}}}{g_{2_\mathrm{b}}}\right),
\label{eq:doke}
\end{equation}
where $W_q = 13.7\,\mathrm{eV}$, and $Q^c_{S1}$ and $Q^c_{S2_\mathrm{b}}$ are the corrected $S1$ and bottom $S2$ charges in photoelectrons, respectively.
The detector parameters ($g_1$, $g_{2_\mathrm{b}}$) are prefitted using the monoenergy ER peaks of  
\,\nclm{Kr-83} (41.5~keV), \nclm{Xe-131} (163.9~keV), and \nclm{Xe-129} (236.2~keV). Because of variations in the operation conditions such as the liquid levels, the values of ($g_1$, $g_{2_\mathrm{b}}$) in subsets of Run0 and Run1 are related by constant factors, which can be tracked and boot-strapped using {\it in situ} $\alpha$ peaks from radon decay. The final values of ($g_1$, $g_{2_\mathrm{b}}$) 
used for the DM analysis is derived from the best fit to the continuous ER 
and NR 
calibration data~\cite{Luo2024_P4SignalModel}.
The mean values in Run0 and Run1 are (0.100$\pm$0.005, 4.1$\pm$0.4) and (0.091$\pm$0.004, 5.0$\pm$0.5).

Comprehensive signal response model: 
To model the ER and NR signal responses in the detector, as well as to optimize selection criteria, several low-energy calibrations are carried out with radioactive sources. The ER response is calibrated using the $\beta$-decay progenies of \ncl{Rn-220} and \ncl{Rn-222}, which are injected through one of the circulation loops~\cite{Meng2021_P4First}. Neutrons produced by a deuterium-deuterium (DD) neutron generator or an \ncl{Am-241}Be source are employed to calibrate the NR response.
A comprehensive signal model of the PandaX-4T detector is developed by combining a scintillation and ionization production model based on the NESTv2 parameterization~\cite{NESTv2}, detector parameters from the previous paragraph, signal efficiencies, and a waveform simulation~\cite{Li2024_WfSim} which reassembles PMT waveforms from the real data. This allows us to apply identical selection cuts to the simulation and real data to assess efficiencies. The parameters in the signal model are derived from a simultaneous fit of all calibration data~\cite{Luo2024_P4SignalModel}.
The goodness-of-fit (GoF) indicates that the model can characterize the response of our detector well.



The following major background compositions are considered, also summarized in Table~\ref{table:bg-summary}. All background components are reestimated using modified data processing and selection efficiencies, and our updated Run0 estimates generally agree with Ref.~\cite{Meng2021_P4First} within uncertainties, except for the neutron and AC background (see later).

One of the top ER background contributions comes from $\beta$ decays of \,\ncl{Pb-214} and \ncl{Pb-212}, the decay progenies of \ncl{Rn-222} and \ncl{Rn-220}, respectively.  
The average decay rate of \ncl{Rn-222} is measured with its decay $\alpha$'s at a level of $7.1 \pm 0.2$ $\mu$Bq/kg in Run0, and increases to an average of $8.7 \pm 0.3$ $\mu$Bq/kg in Run1.
The level of \ncl{Pb-214} is depleted from \ncl{Rn-222} due to the recirculation flow and electric field in the TPC~\cite{Ma2020_RnDepletion}.
We take the low-energy end of the best-fit \ncl{Pb-214} spectrum from Ref.~\cite{Yan2024_Xe134DBD} for Run0, then scale its contribution to Run1 according to the measured \ncl{Rn-222} $\alpha$ rate.
Similar procedures are adopted for $\beta$'s from \,\ncl{Pb-212}.
Because of the short half-life of \ncl{Rn-220} ($55$~s) and the long half-life of \ncl{Pb-212} (10.6~hr), \ncl{Rn-220} $\alpha$ rate does not reflect the internal activity of \,\ncl{Pb-212}.
Instead, the level of the daughter \ncl{Po-212} $\alpha$ decays is taken as a proxy, combined with the depletion factor measured in a \ncl{Rn-220} injection, leading to a \ncl{Pb-212} level of $0.3 \pm 0.1$ $\mu$Bq/kg for both Run0 and Run1.

The background from \ncl{Kr-85} $\beta$-decay is estimated through the correlated emission of $\beta$-$\gamma$ via the metastable state \nclm{Rb-85} (514 keV) with a branching ratio of $0.44\%$. 
Average Kr/Xe ratios in Run0 and Run1 are $0.52 \pm 0.27$~ppt and $0.94 \pm 0.28$~ppt respectively, under the assumption of a $2\times10^{-11}$ isotropic concentration of \ncl{Kr-85}~\cite{Collon2004_forKr}, leading to an expected background of $80 \pm 40$ and $289 \pm 88$ events within the DM detection window in Run0 and Run1, respectively.

The estimation of neutrino-electron elastic scattering is dominated by solar neutrinos such as $pp$ and \ncl{Be-7} neutrinos, which are adopted from Ref.~\cite{Lu2024_P4Solarpp}, which incorporates neutrino flux from the standard solar model, three-flavor neutrino oscillation, and the standard model neutrino-electron scattering with xenon atomic effects~\cite{Chen2017_AtomicEffectSolarNu}.
The two-neutrino double-$\beta$ decay of \ncl{Xe-136} is calculated based on the half-life measurement from our recent study~\cite{Si2022_Xe136DBD}. Additionally, the double electron capture from the $LL$-shell of \ncl{Xe-124} from XENON~\cite{Aprile2022_Xe124136} is incorporated as a single Gaussian peak at 10~keV. The contributions of $LM$- and $LN$-shell double electron capture are minor and thus not considered in this analysis.

The radioactivity of detector materials, mainly stemming from the PMTs and stainless steel vessels, is assessed using a \Geantbased Monte Carlo (MC) simulation~\cite{Chen2021_BambooMC}.
The data match the simulations in the energy and position distributions for high-energy $\gamma$ rays. The total contribution to background is $147 \pm 12$ events.

The unstable \ncl{Xe-127}, produced through the capture of cosmogenically generated neutrons, decays via electron capture. The $L$-shell x ray with an energy of 5.2~keV contributes to the ER background. In this analysis, the rate in Run1 is estimated from that obtained in Run0~\cite{Meng2021_P4First} using the decay half-life of 36.3~days. The contribution in Run1 is almost negligible. The total background is $7.7 \pm 0.4$ events.

To allow an approximate online assessment of the tritium background, we perform one-dimensional spectral fits only based on $S1$ using a set of relaxed cuts, while keeping $S2$ data blinded. In the final DM data fit, tritium levels from both runs are again left floating. One observes a significant decrease of tritium level in Run1, from $2.04 \pm 0.25$/tonne/day at the end of Run0 to $0.26 \pm 0.07$/tonne/day in Run1. These findings are consistent with an independent analysis with one-dimensional energy spectrum fit~\cite{Zeng2024_P4ERPhys}.

The low-energy neutron background originating from detector materials is evaluated as the weighted average of three different estimates~\cite{Huang2022_NeutronBG}, by direct MC prediction, by scaling the measured MS neutron events, and by scaling the measured high-energy xenon-capture $\gamma$ rays.
The contributions of MS neutrons that satisfy the SS cuts but with only partial ionization energy collected (known as ``neutron-X''~\cite{Huang2022_NeutronBG}) are also taken into account in all three estimates.
Compared with Ref.~\cite{Meng2021_P4First}, major updates in the neutron background estimates include:
(1) an update of the NR signal response model (mentioned earlier) from raw energy deposition all the way to the PMT waveform level, thereby more accurate SS tagging~\cite{Luo2024_P4SignalModel};
(2) a reduced FV (see later);
(3) an improved input material radioactivities according to spectral fit performed at high energy~\cite{Si2022_Xe136DBD};
(4) an update of the identified MS events in the data;
(5) an improved identification of high-energy neutron capture $\gamma$ events in the data.
The combined effect is that the neutron background is reduced by $44\%$ in comparison to Ref.~\cite{Meng2021_P4First}, with a total of $1.7 \pm 0.9$ events for the two runs.

Solar \,\ncl{B-8} neutrinos can produce coherent elastic scattering (CE$\nu$NS)~\cite{Ruppin2014_NeutrinoBg}, thereby contributing to the NR background in the low-energy region.
Our new estimate of 0.31 events for Run0 is about half of that in Ref.~\cite{Meng2021_P4First}, due to the update of the signal response model and a smaller FV, yet still consistent within the previously assigned uncertainty.
The total contribution of \,\ncl{B-8} CE$\nu$NS is $1.0 \pm 0.3$ events in Run0 and Run1 combined.

The surface background comes from the low-energy $\beta$-decay events due to radon plate-out on the PTFE panels. The $S2$ of these events are significantly suppressed due to distorted electric field lines close to the PTFE surface. A data-driven background model is established using \ncl{Po-210} $\alpha$ surface events, whose $S2$s are similarly suppressed. Within the optimized FV selection, the total surface background is estimated to be $0.26 \pm 0.12$ events.

Unphysical accidental coincidence (AC) backgrounds are estimated similarly as in Ref.~\cite{Abdukerim2022_AC}.
The $S1$ ($S2$) identified in a fixed 2-ms event window without any pairable $S2$ ($S1$) is defined as an isolated $S1$ (isolated $S2$).
In Run0 and Run1, the rates of isolated $S1$ (isolated $S2$) are $13.3$~Hz ($0.125$~Hz) and $14.4$~Hz ($0.146$~Hz), respectively, under a set of loose selection cuts.
The AC rate is estimated by manually pairing randomly scrambled isolated-$S1$ and isolated-$S2$ pulses, and then applying the full data selection procedure. 
The preunblinding AC backgrounds were estimated as $5.6 \pm 1.7$ and $7.9 \pm 2.4$ events for Run0 and Run1, respectively~\footnote{These values are updated after unblinding, see later.}.
In the two runs, numbers of AC background do not quite scale with exposures primarily due to different cut efficiencies.
Compared to Ref.~\cite{Meng2021_P4First}, a higher AC background level arises from the updated signal reconstruction procedures and a set of new selection cuts with higher signal efficiency.
The selection cuts are determined by optimizing the figure-of-merit (FoM) based on the signal and AC background below the NR median region.

\begin{table*}[t]
\caption{\label{table:bg-summary}
    Expected background contributions to DM candidates for Run0 and Run1 with their uncertainties.
    ``Other ER (data)'' combines contributions from radon progenies, krypton, backgrounds related to detector materials, elastic solar neutrinos-electron scattering (E$\nu$ES), and \ncl{Xe-136}.
    It is independently derived for each dataset using the normalization obtained from 20 to 30~keV.
    Tritium contributions are determined through unconstrained fit.
    The neutron, \ncl{B-8} CE$\nu$NS, surface, and accidental backgrounds are assumed to be constant for each run.
    The below-NR-median neutron background is more than $50\%$ of the total neutron background, due to contribution from ``neutron-X'' which leads to a smaller $S2/S1$.
    The best-fit values from a background-only fit are also shown.
}

\renewcommand{\arraystretch}{1.25}
\centering
\begin{ruledtabular}
\begin{tabular}{ccccccc}
                            &       Run0       &       Run1        &       Total       &   Below NR median    &     Best fit      \\ 
\hline\noalign{\smallskip}
\ncl{Pb-214}                &  $281 \pm 13$    &  $675 \pm 35$     &  $956 \pm 38$ & $ 3.6^{+0.9}_{-0.7}$  &  $\cdots$    \\
\ncl{Pb-212}                &  $ 49 \pm 13$    &  $ 97 \pm 25$     &  $146 \pm 30$ & $ 0.6^{+0.2}_{-0.2}$  &  $\cdots$       \\
\ncl{Kr-85}                 &  $ 80 \pm 40$    &  $289 \pm 88$     &  $369 \pm 96$ & $ 1.4^{+0.5}_{-0.5}$  &  $\cdots$     \\
Material ER                 &  $ 42 \pm  5$    &  $105 \pm 11$     &  $147 \pm 12$ & $ 0.6^{+0.2}_{-0.1}$  &  $\cdots$ \\
E$\nu$ES                    &  $ 38 \pm  4$    &  $ 74 \pm  7$     &  $ 111 \pm  8$ & $ 0.4^{+0.1}_{-0.1}$  &  $\cdots$ \\
\ncl{Xe-136}                &  $ 28 \pm  1$    &  $ 59 \pm  3$     &  $ 87 \pm  3$ & $ 0.2^{+0.1}_{-0.1}$  &  $\cdots$ \\
\midrule\noalign{\smallskip}
Other ER (data)             &  $504 \pm 16$    & $1226 \pm 28$     & $1730 \pm 32$ & $ 6.4^{+1.7}_{-1.2}$ & $1767\pm 48$         \\
CH$_3$T                     &  $556 \pm 33$    &  $114 \pm 33$     &  $670 \pm 47$ &  $5.2^{+1.2}_{-1.1}$ & $677\pm 44$ \\
\ncl{Xe-127}                &  $7.7 \pm 0.4$   &  $ 0.0 \pm 0.0$   &  $ 7.7 \pm 0.4$ &  $0.10^{+0.02}_{-0.02}$ & $7.7\pm 0.4$  \\
\ncl{Xe-124}                &  $2.3 \pm 0.6$   &  $ 4.1 \pm 1.1$   &  $ 6.3 \pm 1.7$ & $ 0.03^{+0.01}_{-0.01}$ & $6.2\pm 1.7$  \\
Neutron                     &  $0.6 \pm 0.3$   &  $ 1.1 \pm 0.6$   &  $ 1.7 \pm 0.9$ & $ 1.0^{+0.5}_{-0.5}$ & $1.6\pm 0.8$  \\
\ncl{B-8} CE$\nu$NS         &  $0.3 \pm 0.1$ &  $ 0.7 \pm 0.2$ &  $ 1.0 \pm 0.3$ & $ 1.0^{+0.3}_{-0.3}$ & $1.1 \pm 0.3$  \\
Surface                     &  $0.09 \pm 0.06$ &  $ 0.17 \pm 0.11$ &  $ 0.26 \pm 0.12$ & $ 0.26^{+0.12}_{-0.12}$ & $0.25\pm 0.11$ \\
Accidental                  &  $ 11 \pm 3$     &  $13 \pm 4$       &  $24 \pm 5$ & $ 6.4^{+1.4}_{-1.4}$ &$26\pm 5$\\
Sum                         & $1082 \pm 37$    & $1356 \pm 43$     & $2439 \pm 43$& $ 20.5^{+2.5}_{-2.2}$ & $2487 \pm 94$ \\ 
\midrule\noalign{\smallskip}
Observed                    &     $1117$       &      $1373$       &       $2490$      &     $24$      &    $\cdots$      \\
\bottomrule
\end{tabular}
\end{ruledtabular}
\end{table*}


\begin{figure}[!htbp]
    \centering
    \includegraphics[width=0.48\textwidth]{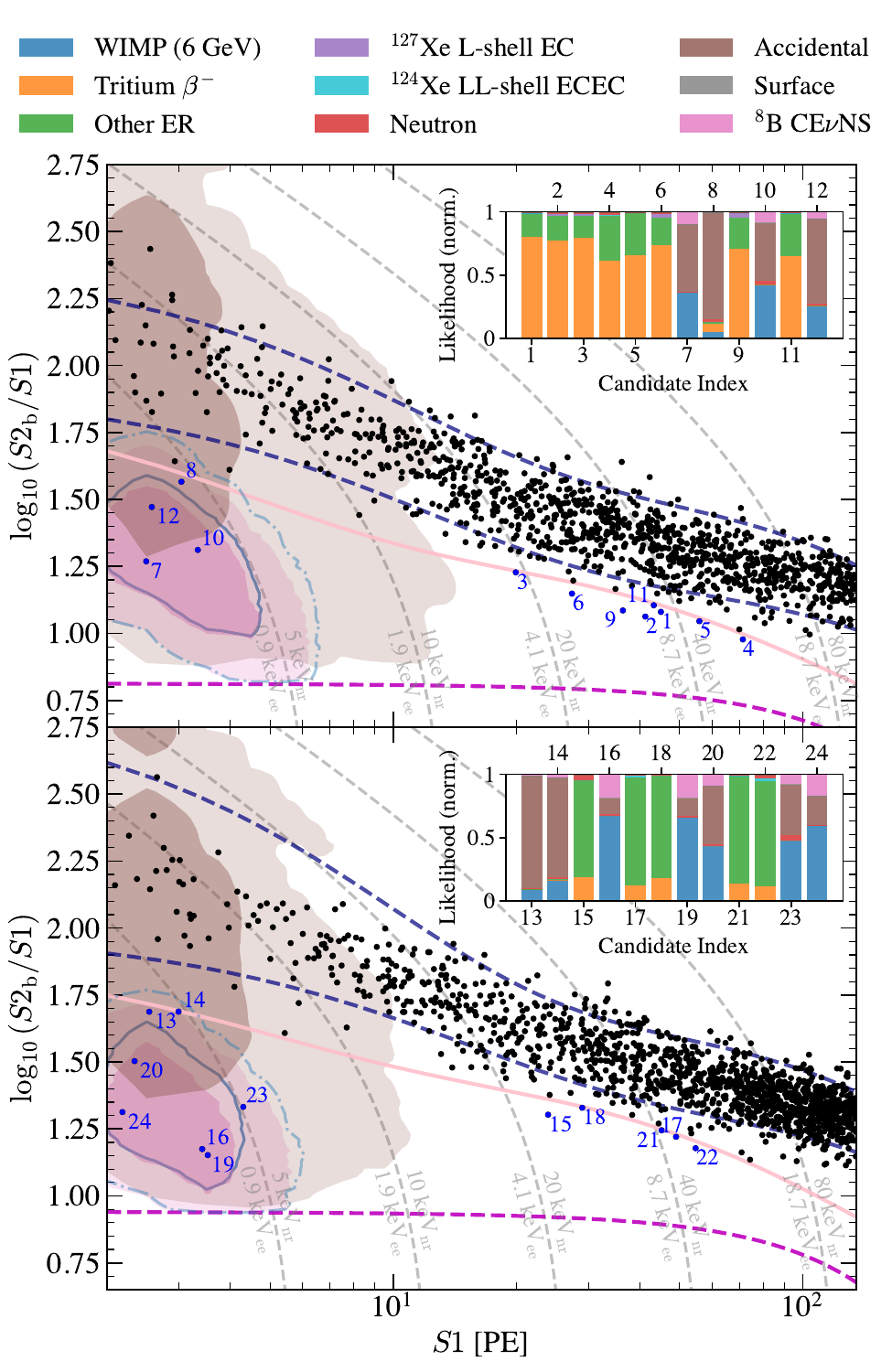}
    \caption{Distributions of the Run0 (top) and Run1 (bottom) final DM candidates in $\log_{10}(S2_{\rm b}/S1)$ vs $S1$, overlaid with $5\%$ and $95\%$ ER quantiles (dashed navy blue), NR median curves (solid pink), $99.5\%$ NR acceptance boundaries (dashed violet), and the equal ER/NR energy curves (dashed gray).
    The NR band is derived from a combination of AmBe and DD NR events.
    Events located below the NR median curves are highlighted by blue dots (numbered according to the order of calendar time), with their normalized likelihood values for signal and background components (see legends) stacked in the insets.
    The $1\sigma$ (dark) and $2\sigma$ (light) contours are also overlaid: AC (brown), \ncl{B-8} CE$\nu$NS (pink), and $6$~\GeV DM (blue solid and dash-dotted line).
    }
    \label{fig:event-dist-band}
\end{figure}

\begin{figure}[!htbp]
    \centering
    \begin{subfigure}{0.42\textwidth}
    \caption{$Y$ vs $X$}
    \includegraphics[width=0.95\textwidth]{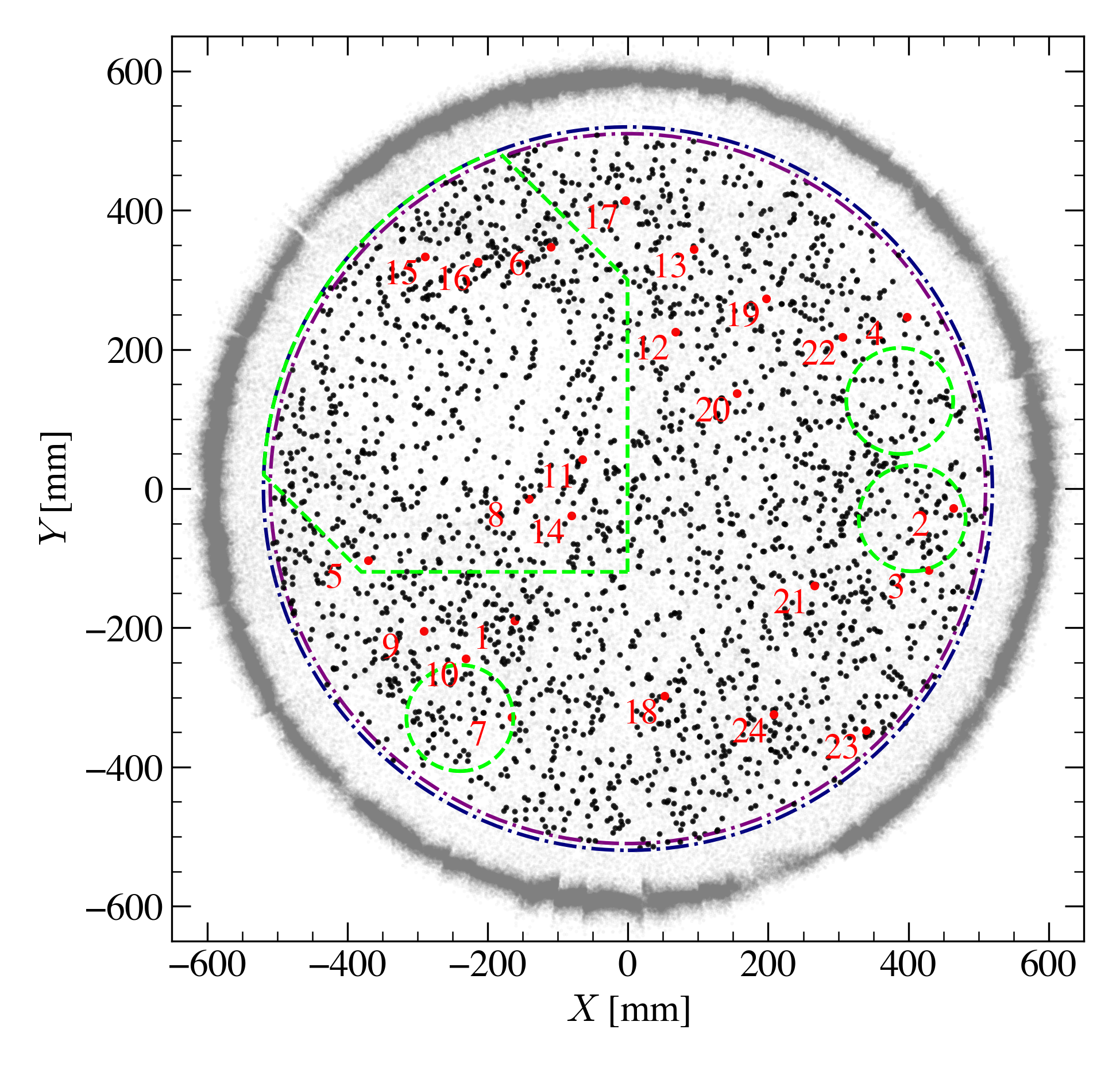}
    \label{fig:pos_y_vs_x}
    \end{subfigure}
    \begin{subfigure}{0.42\textwidth}
    \caption{$Z$ vs $R^{2}$}
    \includegraphics[width=0.95\textwidth]{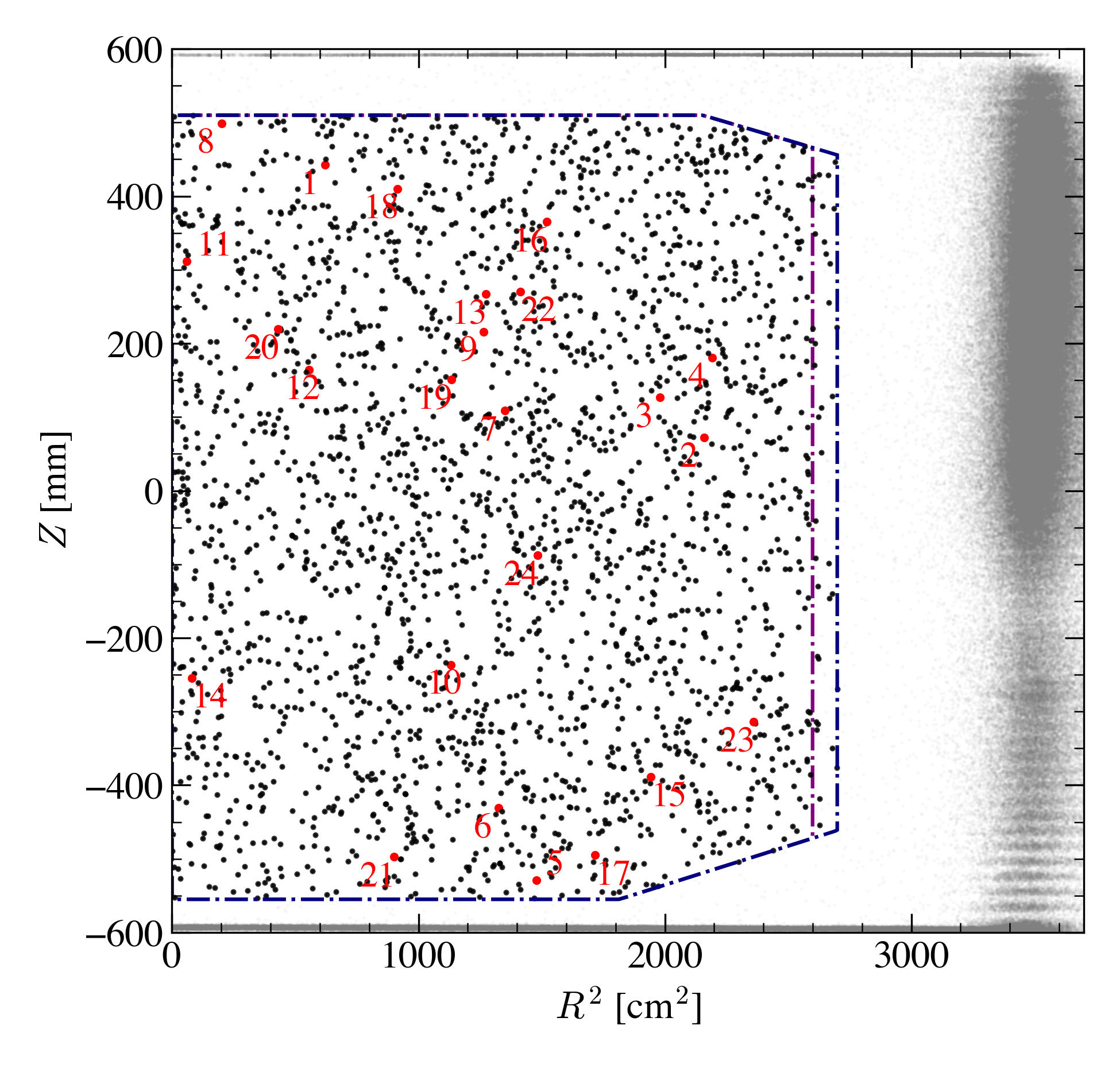}
    \label{fig:pos_z_vs_r2}
    \end{subfigure}
    \caption{
    Spatial distributions of the final dark matter candidates in (a) $Y$ vs $X$, and (b) $Z$ vs $R^2$, overlaid with FV for Run0 (purple dashed line) and Run1 (blue dashed line). In (a), areas enclosed by the green dashed lines indicate the regions affected by malfunctioned top PMTs in Run1.
    Events located below the NR median curves are indicated in red dots, with the same numbering scheme as in Fig.~\ref{fig:event-dist-band}.}
    \label{fig:event-dist-pos}
\end{figure}


The FV in Run0 and Run1 is optimized separately, utilizing a FoM based on signal efficiency and background levels~\cite{Cowan2012_DiscoverySF}.
The final fiducial masses are $2.38 \pm 0.04$~tonne (Run0) and $2.48 \pm 0.05$~tonne (Run1).
The uncertainties are derived from the nonuniformity of \nclm{Kr-83} and \,\ncl{Rn-220/222} calibration events, therefore encompassing systematics arising from malfunctioned top PMTs.
The total accumulated exposure is $1.54$~tonne$\cdot$year, with 0.54~tonne$\cdot$year from Run0 and 1.00~tonne$\cdot$year from Run1.

The unblinding of science data was performed with a three-step procedure, by sequentially unblinding about $10\%$, $50\%$, then the entire dataset from both runs, and performing sanity checks on the data selections.
No adjustments were made to the data selection. 
After fully unblinding the data, we realized that our AC background was estimated too low by about a factor of 2, due to an inconsistent acceptance cut on the scrambled data.
Instead of making postunblinding tightening to the cuts, we updated the nominal AC predictions (Table~\ref{table:bg-summary}), which were then fed into the likelihood fits.
As a result, the overall AC contribution in this analysis is increased from Ref.~\cite{Meng2021_P4First}, which weakened the sensitivity at low-mass ranges, but the impact is minor to our DM sensitivity for a DM mass above $40$~\GeV. 

After unblinding, a total of $2490$ final candidate events are identified in Run0 and Run1 within the FV and DM ROI selection window.
As presented in Fig.~\ref{fig:event-dist-band}, twelve (Run0) and twelve (Run1) events are found below the NR median curves in $\log_{10}(S2_{\rm b}/S1)$ vs $S1$.
These events are uniformly distributed inside the TPC (Fig.~\ref{fig:event-dist-pos}) and in time. 
Compared with the six events found below the NR median in the previous Run0 analysis~\cite{Meng2021_P4First}, five of them (no. 2, no. 3, no. 5, no. 9, no. 11) are the same, 
and one event is removed by the newly optimized FV and the afterglow cut. All seven newly observed events are near the boundaries of one or two selection cuts: events no. 4, no. 6, no. 7, no. 8, no. 10, and no. 12 do not satisfy previous afterglow or several previous quality cuts, and event no. 1 was above the previous NR median curve.

A standard unbinned likelihood function~\cite{Gu2022_P4NRAbsp, Ning2023_EFT} in [$S1$, $\log_{10}(S2_{\rm b}/S1)$] two-dimensional space is constructed, with an update of nuisance parameters from the combined-fit signal response model utilizing all available low-energy ER and NR calibration data~\cite{Luo2024_P4SignalModel}.
Parameters of interest consist of the amount of DM and background events, and nuisance parameters contain systematics of data selection efficiency, signal response model parameters, as well as background levels.
The likelihood fit takes background and nuisance parameters from Table~\ref{table:bg-summary} as inputs.
Because of their similar distributions in $S1$ and $S2$, we combine the continuous ER components (\,\ncl{Rn-220/222}, \ncl{Kr-85}, solar neutrino, \ncl{Xe-136}, and detector materials) with their nominal contribution in Table~\ref{table:bg-summary} into a single ``Other ER'' component.
The overall normalization is independently determined from the data within 20 to 30 keV, which agrees with the sum of individual components but with better statistical uncertainty.
For each other constrained background component, there is only one nuisance parameter for both runs.
The tritium levels in both runs are set floating. 
In general, background-only fit results agree with preunblinding estimates (Table~\ref{table:bg-summary}). 
For those events below the NR median curves, based on the likelihood composition of each event in Fig.~\ref{fig:event-dist-band}, events with small $S1$ are dominated by AC background and large $S1$ events are ER leaks from tritium in Run0 and ``Other ER'' in Run1.

\begin{figure}[!htbp]
    \centering
    \includegraphics[width=0.48\textwidth]{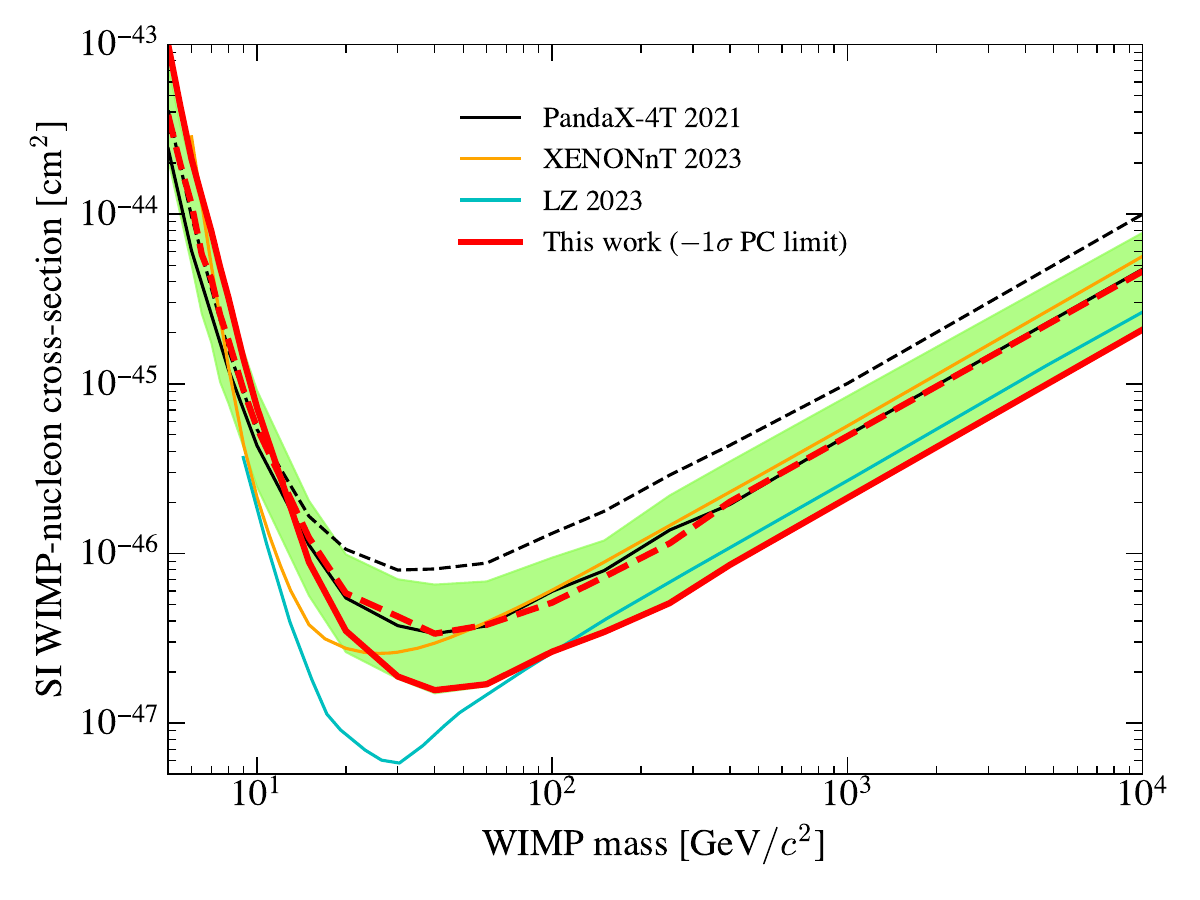}
    \caption{The 90\% CL upper limit of SI DM-nucleon elastic cross section vs $m_{\chi}$ from this work (red), overlaid with that from the LZ 2023~\cite{Aalbers2023_LZFirst} (cyan), XENONnT 2023~\cite{Aprile2023_nTFirst} (orange, extended to 10~\TeV with a $1 / m_{\chi}$ scaling, $-1\sigma$ power constrained) and \PX 2021~\cite{Meng2021_P4First} (black).
    The shaded green region represents the $\pm1\sigma$ sensitivity band of this work, overlaid with its own median curve (red dashed line) and the median sensitivity of PandaX-4T Run0 (black dashed line).
    }
    \label{fig:limit}
\end{figure}

The DM search is carried out for a mass range between $5$~\GeV and $10$~\TeV. The standard conventions for direct dark matter searches~\cite{Baxter2021_ReportResWhitePaper} are adopted, using a profile likelihood ratio method to derive statistical interpretation~\cite{Cowan:2010js}.
The GoF $p$ value under the background-only hypothesis is 0.64, indicating a good fit to the observed data.
For low DM mass below $10$~\GeV, the number of observed events is slightly higher than the expected background.
The largest local significance is found at $6$~\GeV with a value of $1.8\sigma$ (best fit signal of 4.3 events), but reduces to $1.2\sigma$ (global $p$ value of $0.11$) after the look-elsewhere effect correction~\cite{Gross2010_LEE}.
On the other hand, a downward fluctuation is observed for DM masses above $15$~\GeV due to fewer events appearing within the core of the DM signal region.
The derived 90\% confidence level (CL) upper limits (without power constraint~\cite{Cowan:2011an}) on the spin-independent (SI) DM-nucleon cross section is shown in Fig.~\ref{fig:limit}. The limit is slightly outside the $+1\sigma$ sensitivity band at the low mass end, and approaches $-1\sigma$ when the DM mass increases to above 15~\GeV, consistent with the data features discussed above. 
The lowest excluded cross section of $1.6 \times 10^{-47}~\mathrm{cm}^2$ occurs at a DM mass of $40$~\GeV, and the median sensitivity is $3 \times 10^{-47}~\mathrm{cm}^2$ at this DM mass~\footnote{See Supplemental Material at \url{http://link.aps.org/supplemental/10.1103/PhysRevLett.134.011805}, which includes Refs.~\cite{Klos2013_ChiralEFT, Amole2019_PICOSDp, atlas2018_2HDMa_propo, Huang2022_P4SD, atlas2023_2HDMa, Li2018_DMSimplified, Li2019_DMSimplifiedLoopEffects}, for results on alternative models and more information on the statistic test}.
At the high mass region, our limits are very close to the $-1\sigma$ sensitivity boundary.
To be consistent with previous treatment~\cite{Meng2021_P4First}, we choose to report them with power constraints.
In comparison to earlier results from XENONnT~\cite{Aprile2023_nTFirst} and LZ~\cite{Aalbers2023_LZFirst}, we obtain the most stringent upper limit for DM mass larger than $100$~\GeV.

\paragraph{Conclusion—}
In summary, we present the results of a DM search from a combined analysis of data collected during Run0 and Run1 at PandaX-4T, with Run1 data treated in a fully blinded manner, amounting to a cumulative live exposure of $1.54$ tonne$\cdot$year.
No significant event excess is observed above the expected background. Our analysis derives the 90\% CL upper limits on the spin-independent DM-nucleon scattering cross section, with the lowest excluded value of $1.6 \times 10^{-47}~\mathrm{cm}^2$ at 40~\GeV DM mass.
These findings establish a new, stringent constraint at a DM mass exceeding 100~\GeV. During the CJPL construction shutdown, \PX completed upgrades to its PMT, electronics, DAQ, and external water veto systems.
The second science run has now commenced, with ongoing efforts to further suppress background.
We expect that the sensitivity of the DM search will improve by a factor of 2-3 with a complete 6-tonne$\cdot$year exposure.



\paragraph{Acknowledgement—}
This project is supported in part by Grants from National Key R\&D Program of China (Nos. 2023YFA1606200, 2023YFA1606201), National Science Foundation of China (No. 12090060, No. 12090061, No. 12205181, No. 12222505, No. 12325505, No. U23B2070), and by Office of Science and Technology, Shanghai Municipal Government (grant No. 21TQ1400218, No. 22JC1410100, No. 23JC1410200, No. ZJ2023-ZD-003). We are thankful for the support by the Fundamental Research Funds for the Central Universities. We also thank the sponsorship from the Chinese Academy of Sciences Center for Excellence in Particle Physics (CCEPP), Hongwen Foundation in Hong Kong, New Cornerstone Science Foundation, Tencent Foundation in China, and Yangyang Development Fund. Finally, we thank the CJPL administration and the Yalong River Hydropower Development Company Ltd. for indispensable logistical support and other help.

This work is the result of the contributions and efforts of all participating institutes of the PandaX Collaboration, under the leadership of the hosting institute, Shanghai Jiao Tong University.
The collaboration has constructed and operated the PandaX-4T apparatus, and performed the data processing, calibration, and data selections.
J. Liu is the Collaboration Spokesperson.
Z. Bo, Z. Gao, C. Han, J. Li, Y. Luo, Z. Qian, Y. Yun, X. Zeng, M. Zhang, S. Zhang, and Y. Zhou performed the data analysis and hypothesis test under the guidance of Y. Tao.
X. Zeng, M. Zhang, and Y. Zhou mainly studied the data selection.
Z. Bo, Y. Luo, and S. Zhang mainly studied the signal response.
Z. Qian, J. Li, Y. Yun, and Z. Gao mainly studied the background.
C. Han mainly performed the hypothesis test.
The paper draft was prepared by Y. Tao, and extensively edited by J. Liu.
All authors approved the final version of the manuscript.

\vspace{1em}

\begin{figure}[!htbp]
    \centering
    \includegraphics[width=0.48\textwidth]{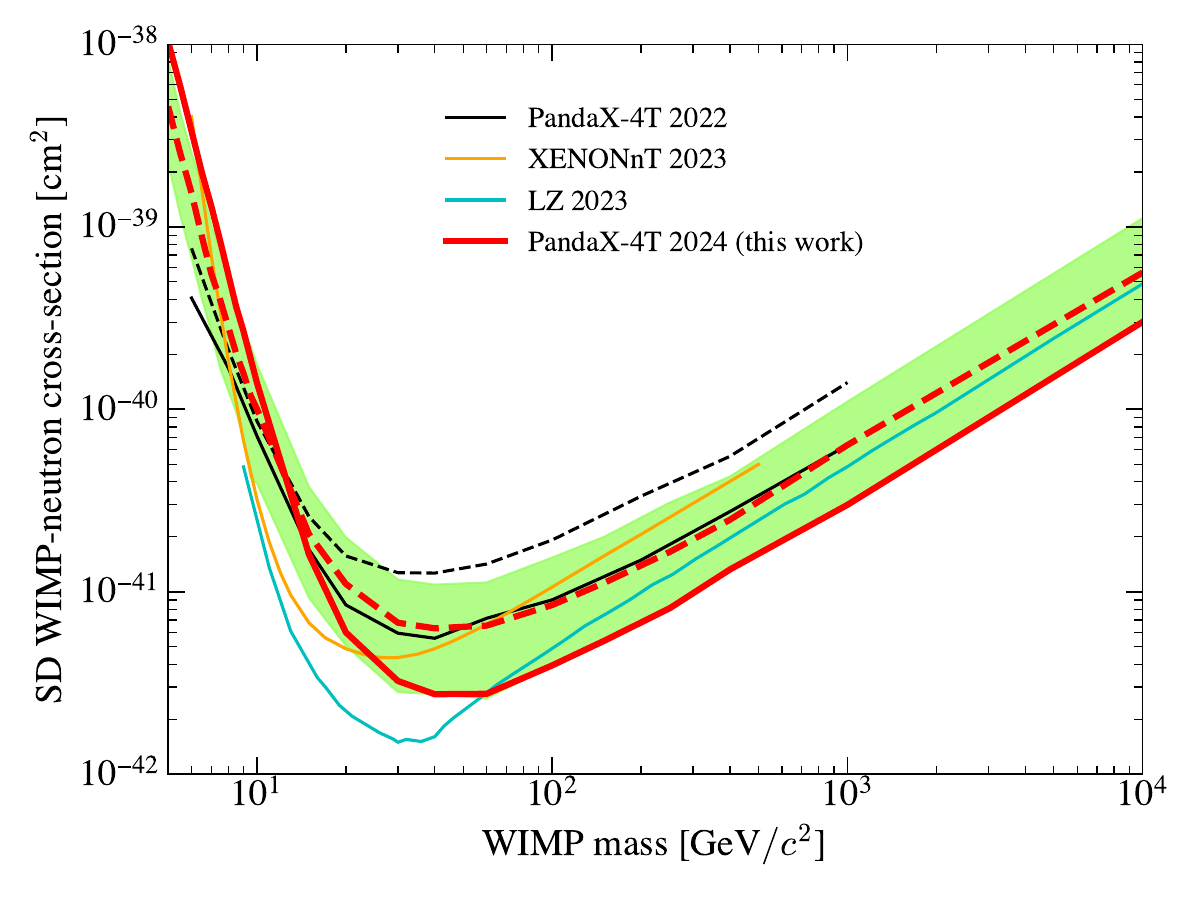}
    \caption{The 90\% CL upper limit from this work vs $m_{\chi}$ for the SD DM-neutron cross-section, overlaid with that from the LZ 2023~\cite{Aalbers2023_LZFirst} (cyan), XENONnT 2023~\cite{Aprile2023_nTFirst} (orange) and \PX 2021~\cite{Meng2021_P4First} (black, with the median sensitivity in dashed line).
    The sensitivity band, corresponding to $\pm1\sigma$, is depicted as a region shaded in green, with its median indicated by a red dashed line.
    }
    \label{fig:sdn_limit}
\end{figure}

\begin{figure}[!htbp]
    \centering
    \includegraphics[width=0.48\textwidth]{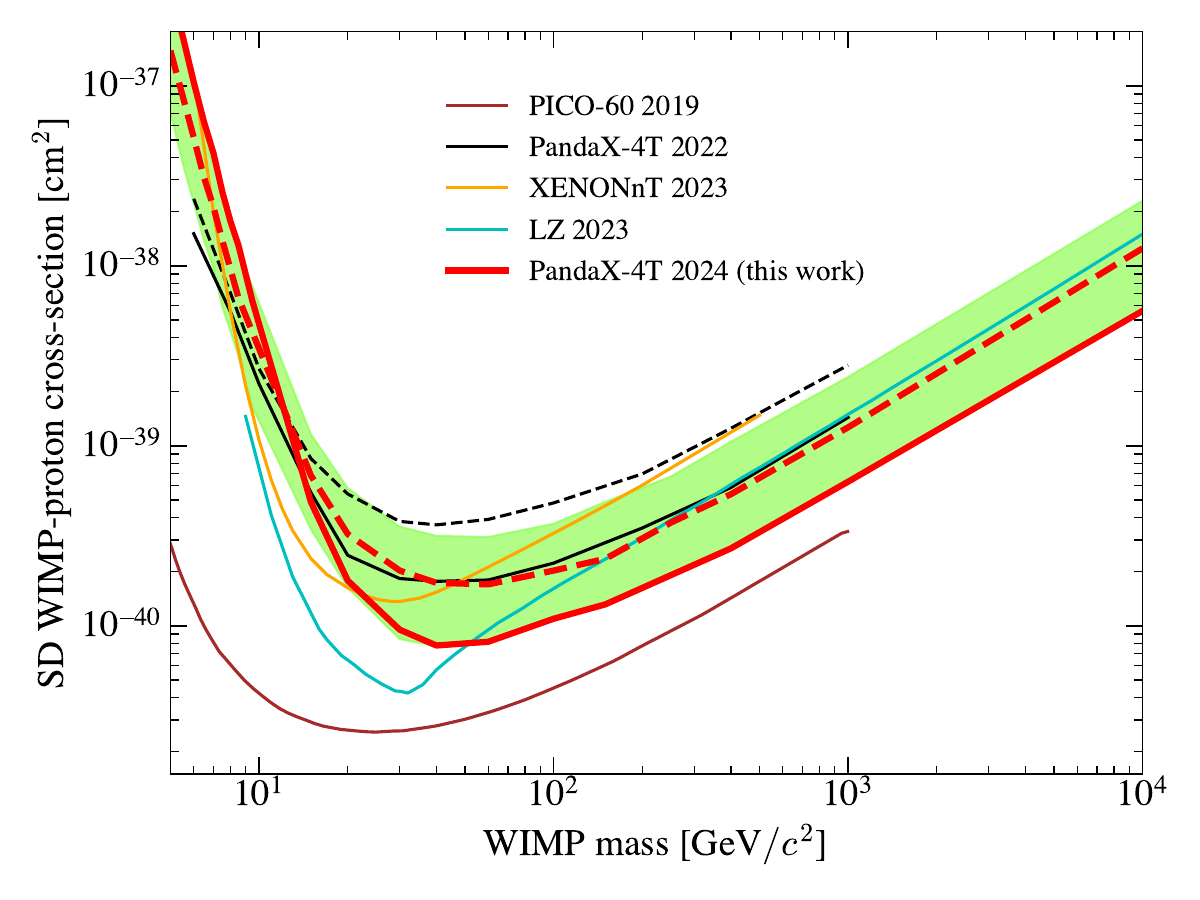}
    \caption{The 90\% CL upper limit from this work vs $m_{\chi}$ for the SD DM-proton cross-section, overlaid with that from the PICO-60 2019~\cite{Amole2019_PICOSDp} (brown), LZ 2023~\cite{Aalbers2023_LZFirst} (cyan), XENONnT 2023~\cite{Aprile2023_nTFirst} (orange) and \PX 2021~\cite{Meng2021_P4First} (black, with the median sensitivity in dashed line).
    The sensitivity band, corresponding to $\pm1\sigma$, is depicted as a region shaded in green, with its median indicated by a red dashed line.
    }
    \label{fig:sdp_limit}
\end{figure}

\paragraph{Appendix—}
Alternative DM models results: Significant abundance of odd-$A$ xenon isotopes have non-zero nuclear spin (26.4\% spin-$1/2$ \,\ncl{Xe-129} and 21.2\% spin-$3/2$ \,\ncl{Xe-131} in natural xenon).
Using the same $1.54$~tonne$\cdot$year physics data of PandaX-4T, we also perform dedicated searches on spin-dependent (SD) WIMP-nucleon interaction, with xenon nucleus spin structure factors for neutron and proton taken from a chiral effective theory calculation~\cite{Klos2013_ChiralEFT}.
The 90\% CL upper limits of ``neutron-only'' and ``proton-only'' cases are shown in Fig.~\ref{fig:sdn_limit} and Fig.~\ref{fig:sdp_limit}, respectively, overlaid with the latest updates of SD-constraints from the xenon-based experiments~\cite{Meng2021_P4First, Aalbers2023_LZFirst, Aprile2023_nTFirst}.

\begin{figure}[!htbp]
    \centering
    \includegraphics[width=0.48\textwidth]{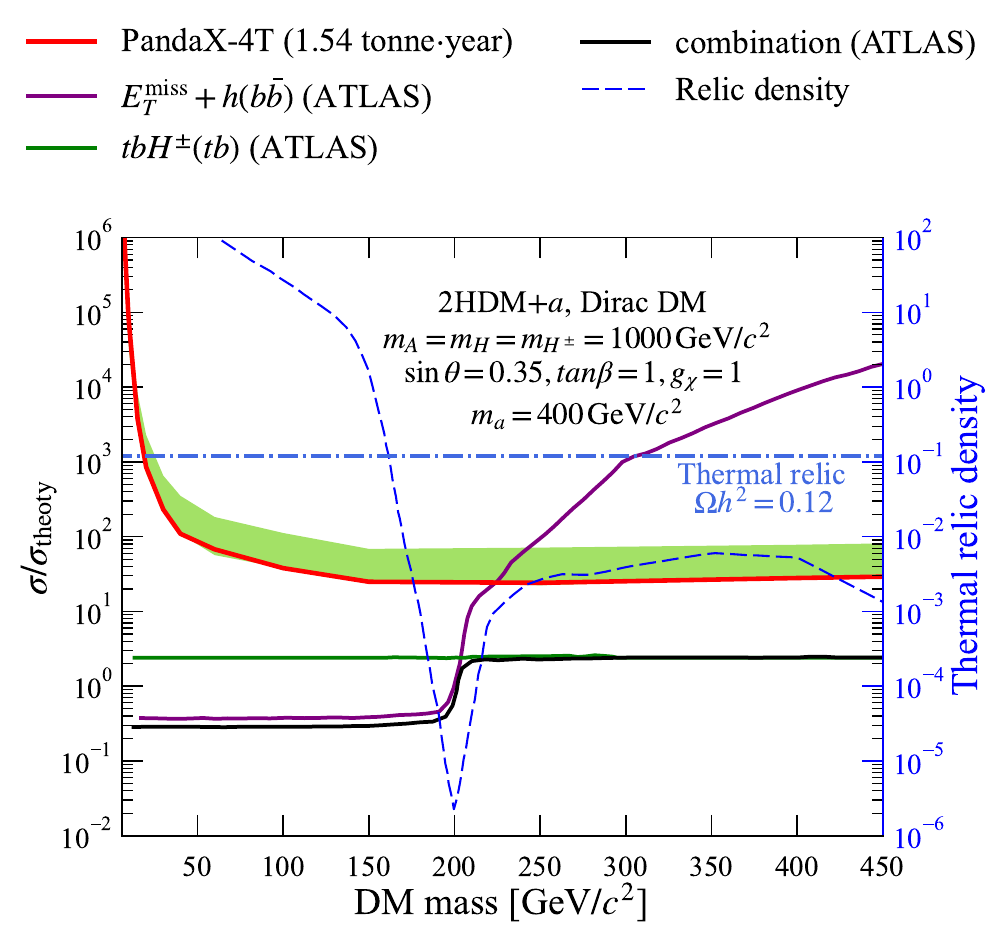}
    \caption{
    The 95\% CL upper limit in the ratio between the excluded cross section to the nominal cross section of the 2HDM$+a$ theory~\cite{atlas2023_2HDMa} from this work vs $m_{\chi}$, overlaid with limits from different channels in ATLAS~\cite{atlas2023_2HDMa}. Limits using missing energy and bottom-quark-pairs $E^\mathrm{miss}_T+h(b\overline{b})$ (purple solid) and final states with top-bottom pairs $tbH^\pm(tb)$ (green solid) dominate in regions where the DM mass is below and above 200~\GeV, respectively. The black solid line shows the result of the combination of all search channels. Under the chosen 2HDM$+a$ model parameters, the predicted dark matter relic density is shown as the blue dashed curve ($\Omega h^2 = 0.12$ indicated as the blue dash-dotted line), with axis on the right.
    }
    \label{fig:2hdma_limit}
\end{figure}

Beyond the generic SI or SD interaction, we also test an ultra-violate complete model with PandaX-4T data, the two-Higgs-doublet model with a pseudo-scalar mediator ($a$) connecting the dark sector to the Standard Model particles.
This model is referred to as the 2HDM$+a$ Model~\cite{atlas2018_2HDMa_propo, Huang2022_P4SD}, which has been widely searched in collider experiments~\cite{atlas2023_2HDMa}. 
In such a model, the SI-interaction is velocity suppressed at the tree level, leading to a more important loop contribution~\cite{Li2018_DMSimplified, Li2019_DMSimplifiedLoopEffects}.
The DM-nucleon elastic scattering formalism in this model is given in Ref.~\cite{atlas2018_2HDMa_propo}.
To directly compare to earlier works, in this letter, the 2HDM$+a$ model parameters are fixed to the values in Ref.~\cite{atlas2023_2HDMa}.
The 95\% CL upper limits on the ratio of the excluded cross section ($\sigma$) to the
nominal cross section of the model ($\sigma_\mathrm{theory}$) are shown.
The results are shown in Fig.~\ref{fig:2hdma_limit}.

\begin{figure}[!htbp]
    \centering
    \includegraphics[width=0.48\textwidth]{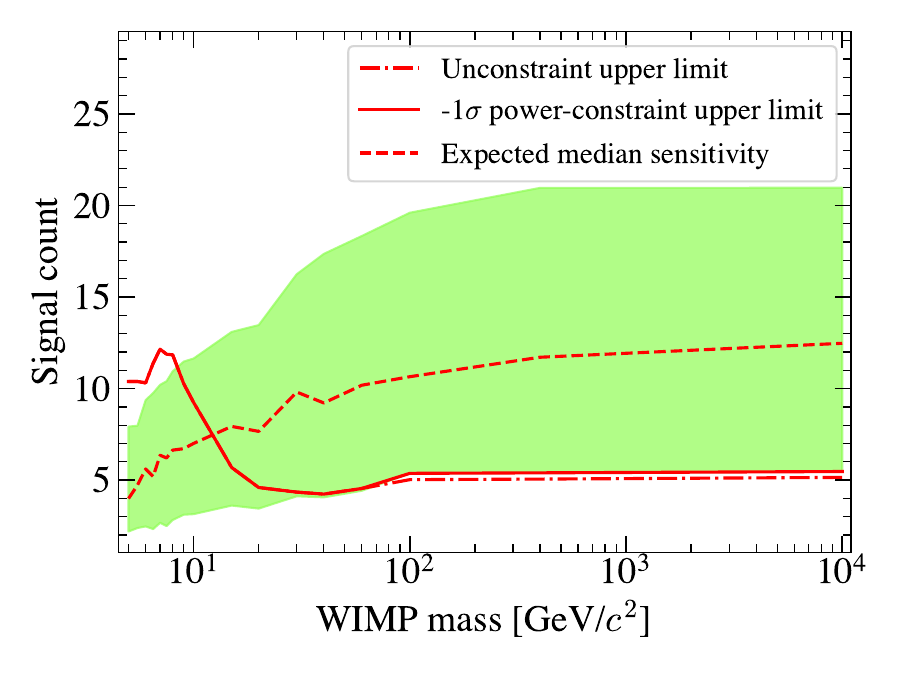}
    \caption{The 90\% CL upper limit with (without) $-1\sigma$ power-constraint of DM counts in solid (dash-dotted) vs $m_{\chi}$ for the combined analysis of PandaX-4T.
    The shaded green region represents the $\pm1\sigma$ sensitivity band, overlaid with its own median curve (red dashed line).
    }
    \label{fig:event_limit}
\end{figure}

Event limits and sensitivity: Figure~\ref{fig:event_limit} presents the Run0+Run1 combined limits in number of DM events.

\begin{figure}[!htbp]
    \centering
    \includegraphics[width=0.48\textwidth]{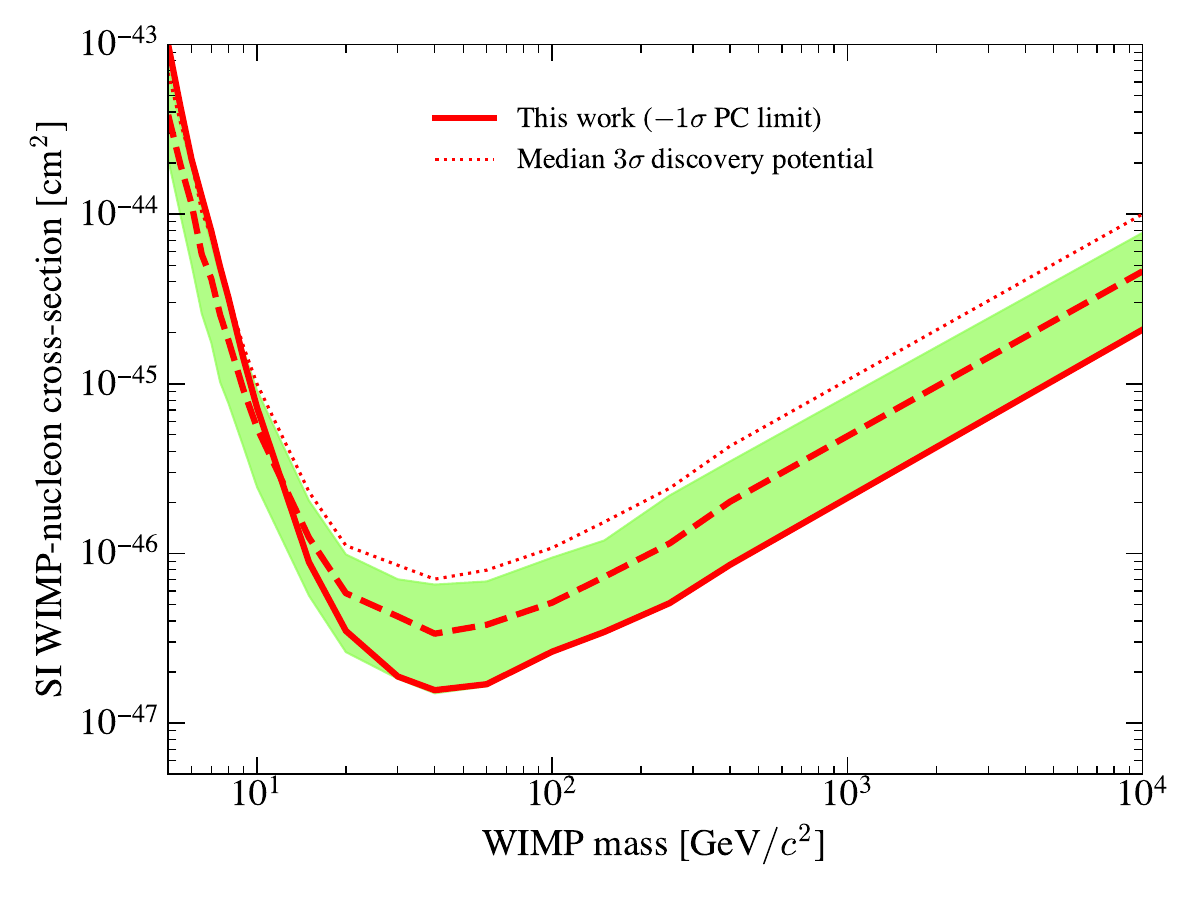}
    \caption{The $3\sigma$ discovery median sensitivity, 
    together with the 90\% CL upper limit with $-1\sigma$ 
    power-constraint and $\pm 1\sigma$ 
    sensitivity band.
    }
    \label{fig:discovery}
\end{figure}

Discovery median sensitivity: Figure~\ref{fig:discovery} presents the $3\sigma$ discovery median sensitivity, which represents the cross-section that yields a $p$-value of $1.4 \times 10^{-3}$~\cite{Baxter2021_ReportResWhitePaper} to be consistent with background under 50\% of trials.

\begin{figure}[!htbp]
    \centering
    \includegraphics[width=0.48\textwidth]{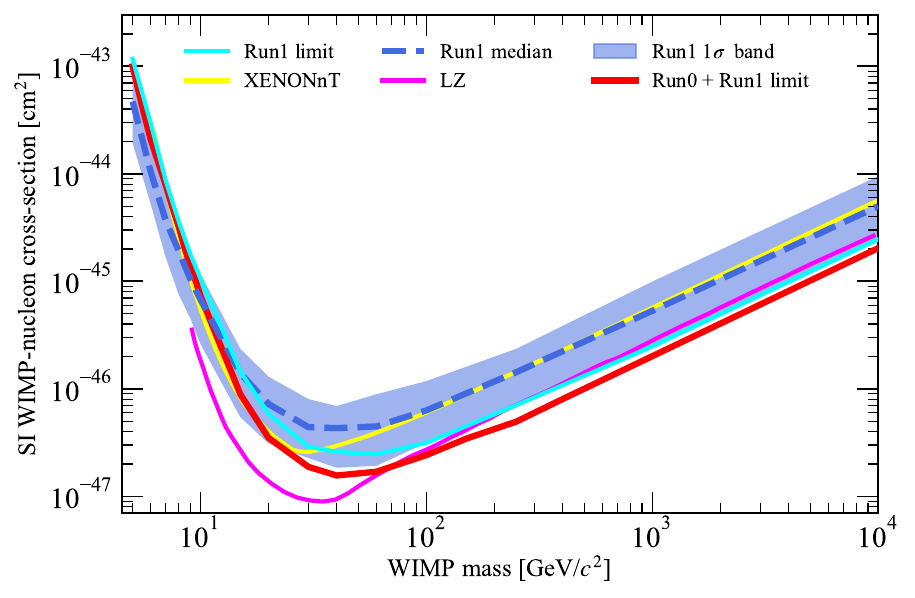}
    \caption{The 90\% CL upper limit of SI DM-nucleon elastic cross section vs $m_{\chi}$ from Run1 data (cyan) and Run0 $+$ Run1 data (this work, red) of PandaX-4T, overlaid with that from the LZ 2023~\cite{Aalbers2023_LZFirst} (magenta), XENONnT 2023~\cite{Aprile2023_nTFirst} (yellow).
    The shaded blue region represents the $\pm1\sigma$ sensitivity band of Run1-only data, overlaid with its own median curve (blue dashed line).
    }
    \label{fig:run1_limit}
\end{figure}

Run1-only limits and sensitivity: To show the improvement introduced by the newly collected Run1 data, we present the 90\% CL upper limit and sensitivity band in Fig.~\ref{fig:run1_limit} using only Run1 data.

\bibliography{apssamp}

\begin{thebibliography}{47}%
\makeatletter
\providecommand \@ifxundefined [1]{%
 \@ifx{#1\undefined}
}%
\providecommand \@ifnum [1]{%
 \ifnum #1\expandafter \@firstoftwo
 \else \expandafter \@secondoftwo
 \fi
}%
\providecommand \@ifx [1]{%
 \ifx #1\expandafter \@firstoftwo
 \else \expandafter \@secondoftwo
 \fi
}%
\providecommand \natexlab [1]{#1}%
\providecommand \enquote  [1]{``#1''}%
\providecommand \bibnamefont  [1]{#1}%
\providecommand \bibfnamefont [1]{#1}%
\providecommand \citenamefont [1]{#1}%
\providecommand \href@noop [0]{\@secondoftwo}%
\providecommand \href [0]{\begingroup \@sanitize@url \@href}%
\providecommand \@href[1]{\@@startlink{#1}\@@href}%
\providecommand \@@href[1]{\endgroup#1\@@endlink}%
\providecommand \@sanitize@url [0]{\catcode `\\12\catcode `\$12\catcode
  `\&12\catcode `\#12\catcode `\^12\catcode `\_12\catcode `\%12\relax}%
\providecommand \@@startlink[1]{}%
\providecommand \@@endlink[0]{}%
\providecommand \url  [0]{\begingroup\@sanitize@url \@url }%
\providecommand \@url [1]{\endgroup\@href {#1}{\urlprefix }}%
\providecommand \urlprefix  [0]{URL }%
\providecommand \Eprint [0]{\href }%
\providecommand \doibase [0]{http://dx.doi.org/}%
\providecommand \selectlanguage [0]{\@gobble}%
\providecommand \bibinfo  [0]{\@secondoftwo}%
\providecommand \bibfield  [0]{\@secondoftwo}%
\providecommand \translation [1]{[#1]}%
\providecommand \BibitemOpen [0]{}%
\providecommand \bibitemStop [0]{}%
\providecommand \bibitemNoStop [0]{.\EOS\space}%
\providecommand \EOS [0]{\spacefactor3000\relax}%
\providecommand \BibitemShut  [1]{\csname bibitem#1\endcsname}%
\let\auto@bib@innerbib\@empty
\bibitem [{\citenamefont {Bertone}\ and\ \citenamefont
  {Hooper}(2018)}]{Bertone:2016nfn}%
  \BibitemOpen
  \bibfield  {author} {\bibinfo {author} {\bibfnamefont {G.}~\bibnamefont
  {Bertone}}\ and\ \bibinfo {author} {\bibfnamefont {D.}~\bibnamefont
  {Hooper}},\ }\href {\doibase 10.1103/RevModPhys.90.045002} {\bibfield
  {journal} {\bibinfo  {journal} {Rev. Mod. Phys.}\ }\textbf {\bibinfo {volume}
  {90}},\ \bibinfo {pages} {045002} (\bibinfo {year} {2018})}\BibitemShut
  {NoStop}%
\bibitem [{\citenamefont {Meng}\ \emph {et~al.}(2021)\citenamefont {Meng} \emph
  {et~al.}}]{Meng2021_P4First}%
  \BibitemOpen
  \bibfield  {author} {\bibinfo {author} {\bibfnamefont {Y.}~\bibnamefont
  {Meng}} \emph {et~al.} (\bibinfo {collaboration} {PandaX Collaboration}),\
  }\href {\doibase 10.1103/PhysRevLett.127.261802} {\bibfield  {journal}
  {\bibinfo  {journal} {Phys. Rev. Lett.}\ }\textbf {\bibinfo {volume} {127}},\
  \bibinfo {pages} {261802} (\bibinfo {year} {2021})}\BibitemShut {NoStop}%
\bibitem [{\citenamefont {Aalbers}\ \emph {et~al.}(2023)\citenamefont {Aalbers}
  \emph {et~al.}}]{Aalbers2023_LZFirst}%
  \BibitemOpen
  \bibfield  {author} {\bibinfo {author} {\bibfnamefont {J.}~\bibnamefont
  {Aalbers}} \emph {et~al.} (\bibinfo {collaboration} {LZ Collaboration}),\
  }\href {\doibase 10.1103/PhysRevLett.131.041002} {\bibfield  {journal}
  {\bibinfo  {journal} {Phys. Rev. Lett.}\ }\textbf {\bibinfo {volume} {131}},\
  \bibinfo {pages} {041002} (\bibinfo {year} {2023})}\BibitemShut {NoStop}%
\bibitem [{\citenamefont {Aprile}\ \emph {et~al.}(2023)\citenamefont {Aprile}
  \emph {et~al.}}]{Aprile2023_nTFirst}%
  \BibitemOpen
  \bibfield  {author} {\bibinfo {author} {\bibfnamefont {E.}~\bibnamefont
  {Aprile}} \emph {et~al.} (\bibinfo {collaboration} {XENON Collaboration}),\
  }\href {\doibase 10.1103/PhysRevLett.131.041003} {\bibfield  {journal}
  {\bibinfo  {journal} {Phys. Rev. Lett.}\ }\textbf {\bibinfo {volume} {131}},\
  \bibinfo {pages} {041003} (\bibinfo {year} {2023})}\BibitemShut {NoStop}%
\bibitem [{\citenamefont {Billard}\ \emph {et~al.}(2014)\citenamefont
  {Billard}, \citenamefont {Figueroa-Feliciano},\ and\ \citenamefont
  {Strigari}}]{Billard2014_NeutrinoFloor}%
  \BibitemOpen
  \bibfield  {author} {\bibinfo {author} {\bibfnamefont {J.}~\bibnamefont
  {Billard}}, \bibinfo {author} {\bibfnamefont {E.}~\bibnamefont
  {Figueroa-Feliciano}}, \ and\ \bibinfo {author} {\bibfnamefont
  {L.}~\bibnamefont {Strigari}},\ }\href {\doibase 10.1103/PhysRevD.89.023524}
  {\bibfield  {journal} {\bibinfo  {journal} {Phys. Rev. D}\ }\textbf {\bibinfo
  {volume} {89}},\ \bibinfo {pages} {023524} (\bibinfo {year}
  {2014})}\BibitemShut {NoStop}%
\bibitem [{\citenamefont {O'Hare}(2021)}]{O’Hare2021_NeutrinoFog}%
  \BibitemOpen
  \bibfield  {author} {\bibinfo {author} {\bibfnamefont {C.~A.~J.}\
  \bibnamefont {O'Hare}},\ }\href {\doibase 10.1103/PhysRevLett.127.251802}
  {\bibfield  {journal} {\bibinfo  {journal} {Phys. Rev. Lett.}\ }\textbf
  {\bibinfo {volume} {127}},\ \bibinfo {pages} {251802} (\bibinfo {year}
  {2021})}\BibitemShut {NoStop}%
\bibitem [{\citenamefont {Zhang}\ \emph {et~al.}(2019)\citenamefont {Zhang}
  \emph {et~al.}}]{Zhang2019_P4Projection}%
  \BibitemOpen
  \bibfield  {author} {\bibinfo {author} {\bibfnamefont {H.}~\bibnamefont
  {Zhang}} \emph {et~al.} (\bibinfo {collaboration} {PandaX Collaboration}),\
  }\href {\doibase 10.1007/s11433-018-9259-0} {\bibfield  {journal} {\bibinfo
  {journal} {Sci. China Phys. Mech. Astron.}\ }\textbf {\bibinfo {volume}
  {62}},\ \bibinfo {pages} {31011} (\bibinfo {year} {2019})}\BibitemShut
  {NoStop}%
\bibitem [{\citenamefont {Kang}\ \emph {et~al.}(2010)\citenamefont {Kang} \emph
  {et~al.}}]{Kang2010}%
  \BibitemOpen
  \bibfield  {author} {\bibinfo {author} {\bibfnamefont {K.~J.}\ \bibnamefont
  {Kang}} \emph {et~al.},\ }\href {\doibase 10.1088/1742-6596/203/1/012028}
  {\bibfield  {journal} {\bibinfo  {journal} {J. Phys. Conf. Ser.}\ }\textbf
  {\bibinfo {volume} {203}},\ \bibinfo {pages} {012028} (\bibinfo {year}
  {2010})}\BibitemShut {NoStop}%
\bibitem [{\citenamefont {Li}\ \emph {et~al.}(2015)\citenamefont {Li},
  \citenamefont {Ji}, \citenamefont {Haxton},\ and\ \citenamefont
  {Wang}}]{Li2014}%
  \BibitemOpen
  \bibfield  {author} {\bibinfo {author} {\bibfnamefont {J.}~\bibnamefont
  {Li}}, \bibinfo {author} {\bibfnamefont {X.}~\bibnamefont {Ji}}, \bibinfo
  {author} {\bibfnamefont {W.}~\bibnamefont {Haxton}}, \ and\ \bibinfo {author}
  {\bibfnamefont {J.~S.~Y.}\ \bibnamefont {Wang}},\ }\href {\doibase
  10.1016/j.phpro.2014.12.055} {\bibfield  {journal} {\bibinfo  {journal}
  {Phys. Proc.}\ }\textbf {\bibinfo {volume} {61}},\ \bibinfo {pages} {576}
  (\bibinfo {year} {2015})}\BibitemShut {NoStop}%
\bibitem [{\citenamefont {Zhao}\ \emph {et~al.}(2021)\citenamefont {Zhao},
  \citenamefont {Cui}, \citenamefont {Ma}, \citenamefont {Fan}, \citenamefont
  {Giboni}, \citenamefont {Zhang}, \citenamefont {Liu},\ and\ \citenamefont
  {Ji}}]{Zhao:2020vxh}%
  \BibitemOpen
  \bibfield  {author} {\bibinfo {author} {\bibfnamefont {L.}~\bibnamefont
  {Zhao}}, \bibinfo {author} {\bibfnamefont {X.}~\bibnamefont {Cui}}, \bibinfo
  {author} {\bibfnamefont {W.}~\bibnamefont {Ma}}, \bibinfo {author}
  {\bibfnamefont {Y.}~\bibnamefont {Fan}}, \bibinfo {author} {\bibfnamefont
  {K.}~\bibnamefont {Giboni}}, \bibinfo {author} {\bibfnamefont
  {T.}~\bibnamefont {Zhang}}, \bibinfo {author} {\bibfnamefont
  {J.}~\bibnamefont {Liu}}, \ and\ \bibinfo {author} {\bibfnamefont
  {X.}~\bibnamefont {Ji}},\ }\href {\doibase 10.1088/1748-0221/16/06/T06007}
  {\bibfield  {journal} {\bibinfo  {journal} {J. Instrum.}\ }\textbf {\bibinfo
  {volume} {16}},\ \bibinfo {pages} {T06007} (\bibinfo {year}
  {2021})}\BibitemShut {NoStop}%
\bibitem [{\citenamefont {Yang}\ \emph {et~al.}(2022)\citenamefont {Yang} \emph
  {et~al.}}]{Yang:2021hnn}%
  \BibitemOpen
  \bibfield  {author} {\bibinfo {author} {\bibfnamefont {J.}~\bibnamefont
  {Yang}} \emph {et~al.},\ }\href {\doibase 10.1088/1748-0221/17/02/T02004}
  {\bibfield  {journal} {\bibinfo  {journal} {J. Instrum.}\ }\textbf {\bibinfo
  {volume} {17}},\ \bibinfo {pages} {T02004} (\bibinfo {year}
  {2022})}\BibitemShut {NoStop}%
\bibitem [{\citenamefont {Gu}\ \emph {et~al.}(2022)\citenamefont {Gu} \emph
  {et~al.}}]{Gu2022_P4NRAbsp}%
  \BibitemOpen
  \bibfield  {author} {\bibinfo {author} {\bibfnamefont {L.}~\bibnamefont {Gu}}
  \emph {et~al.} (\bibinfo {collaboration} {PandaX Collaboration}),\ }\href
  {\doibase 10.1103/PhysRevLett.129.161803} {\bibfield  {journal} {\bibinfo
  {journal} {Phys. Rev. Lett.}\ }\textbf {\bibinfo {volume} {129}},\ \bibinfo
  {pages} {161803} (\bibinfo {year} {2022})}\BibitemShut {NoStop}%
\bibitem [{\citenamefont {Zhang}\ \emph {et~al.}(2022)\citenamefont {Zhang}
  \emph {et~al.}}]{Zhang2022_P4ERAbsp}%
  \BibitemOpen
  \bibfield  {author} {\bibinfo {author} {\bibfnamefont {D.}~\bibnamefont
  {Zhang}} \emph {et~al.} (\bibinfo {collaboration} {PandaX Collaboration}),\
  }\href {\doibase 10.1103/PhysRevLett.129.161804} {\bibfield  {journal}
  {\bibinfo  {journal} {Phys. Rev. Lett.}\ }\textbf {\bibinfo {volume} {129}},\
  \bibinfo {pages} {161804} (\bibinfo {year} {2022})}\BibitemShut {NoStop}%
\bibitem [{\citenamefont {Ning}\ \emph {et~al.}(2023)\citenamefont {Ning} \emph
  {et~al.}}]{Ning2023_EFT}%
  \BibitemOpen
  \bibfield  {author} {\bibinfo {author} {\bibfnamefont {X.}~\bibnamefont
  {Ning}} \emph {et~al.} (\bibinfo {collaboration} {PandaX Collaboration}),\
  }\href {\doibase 10.1038/s41586-023-05982-0} {\bibfield  {journal} {\bibinfo
  {journal} {Nature (London)}\ }\textbf {\bibinfo {volume} {618}},\ \bibinfo
  {pages} {47} (\bibinfo {year} {2023})}\BibitemShut {NoStop}%
\bibitem [{\citenamefont {Cui}\ \emph {et~al.}(2021)\citenamefont {Cui} \emph
  {et~al.}}]{Cui2021_P4Distillation}%
  \BibitemOpen
  \bibfield  {author} {\bibinfo {author} {\bibfnamefont {X.}~\bibnamefont
  {Cui}} \emph {et~al.},\ }\href {\doibase 10.1088/1748-0221/16/07/P07046}
  {\bibfield  {journal} {\bibinfo  {journal} {JINST}\ }\textbf {\bibinfo
  {volume} {16}},\ \bibinfo {pages} {P07046} (\bibinfo {year}
  {2021})}\BibitemShut {NoStop}%
\bibitem [{\citenamefont {Cui}\ \emph {et~al.}(2024)\citenamefont {Cui} \emph
  {et~al.}}]{Cui2024_RnRemove}%
  \BibitemOpen
  \bibfield  {author} {\bibinfo {author} {\bibfnamefont {X.}~\bibnamefont
  {Cui}} \emph {et~al.},\ }\href {\doibase 10.1088/1748-0221/19/07/P07010}
  {\bibfield  {journal} {\bibinfo  {journal} {JINST}\ }\textbf {\bibinfo
  {volume} {19}},\ \bibinfo {pages} {P07010} (\bibinfo {year}
  {2024})}\BibitemShut {NoStop}%
\bibitem [{\citenamefont {Elsied}\ \emph {et~al.}(2015)\citenamefont {Elsied},
  \citenamefont {Giboni},\ and\ \citenamefont {Ji}}]{Elsied:2015ixa}%
  \BibitemOpen
  \bibfield  {author} {\bibinfo {author} {\bibfnamefont {A.~M.~M.}\
  \bibnamefont {Elsied}}, \bibinfo {author} {\bibfnamefont {K.~L.}\
  \bibnamefont {Giboni}}, \ and\ \bibinfo {author} {\bibfnamefont
  {X.}~\bibnamefont {Ji}},\ }\href {\doibase 10.1088/1748-0221/10/01/T01003}
  {\bibfield  {journal} {\bibinfo  {journal} {J. Instrum.}\ }\textbf {\bibinfo
  {volume} {10}},\ \bibinfo {pages} {T01003} (\bibinfo {year}
  {2015})}\BibitemShut {NoStop}%
\bibitem [{\citenamefont {Yan}\ \emph {et~al.}(2024)\citenamefont {Yan} \emph
  {et~al.}}]{Yan2024_Xe134DBD}%
  \BibitemOpen
  \bibfield  {author} {\bibinfo {author} {\bibfnamefont {X.}~\bibnamefont
  {Yan}} \emph {et~al.} (\bibinfo {collaboration} {PandaX Collaboration}),\
  }\href {\doibase 10.1103/PhysRevLett.132.152502} {\bibfield  {journal}
  {\bibinfo  {journal} {Phys. Rev. Lett.}\ }\textbf {\bibinfo {volume} {132}},\
  \bibinfo {pages} {152502} (\bibinfo {year} {2024})}\BibitemShut {NoStop}%
\bibitem [{\citenamefont {Luo}\ \emph {et~al.}(2024)\citenamefont {Luo} \emph
  {et~al.}}]{Luo2024_P4SignalModel}%
  \BibitemOpen
  \bibfield  {author} {\bibinfo {author} {\bibfnamefont {Y.}~\bibnamefont
  {Luo}} \emph {et~al.} (\bibinfo {collaboration} {PandaX Collaboration}),\
  }\href {\doibase 10.1103/PhysRevD.110.023029} {\bibfield  {journal} {\bibinfo
   {journal} {Phys. Rev. D}\ }\textbf {\bibinfo {volume} {110}},\ \bibinfo
  {pages} {023029} (\bibinfo {year} {2024})}\BibitemShut {NoStop}%
\bibitem [{\citenamefont {Li}\ \emph {et~al.}(2024)\citenamefont {Li} \emph
  {et~al.}}]{Li2024_WfSim}%
  \BibitemOpen
  \bibfield  {author} {\bibinfo {author} {\bibfnamefont {J.}~\bibnamefont {Li}}
  \emph {et~al.},\ }\href {\doibase 10.1088/1674-1137/ad380f} {\bibfield
  {journal} {\bibinfo  {journal} {Chin. Phys. C}\ }\textbf {\bibinfo {volume}
  {48}},\ \bibinfo {pages} {073001} (\bibinfo {year} {2024})}\BibitemShut
  {NoStop}%
\bibitem [{\citenamefont {Zhang}\ \emph {et~al.}(2021)\citenamefont {Zhang}
  \emph {et~al.}}]{Zhang2021_PosRecon}%
  \BibitemOpen
  \bibfield  {author} {\bibinfo {author} {\bibfnamefont {D.}~\bibnamefont
  {Zhang}} \emph {et~al.},\ }\href {\doibase 10.1088/1748-0221/16/11/p11040}
  {\bibfield  {journal} {\bibinfo  {journal} {Journal of Instrumentation}\
  }\textbf {\bibinfo {volume} {16}},\ \bibinfo {pages} {P11040} (\bibinfo
  {year} {2021})}\BibitemShut {NoStop}%
\bibitem [{\citenamefont {Szydagis}\ \emph {et~al.}()\citenamefont {Szydagis}
  \emph {et~al.}}]{NESTv2}%
  \BibitemOpen
  \bibfield  {author} {\bibinfo {author} {\bibfnamefont {M.}~\bibnamefont
  {Szydagis}} \emph {et~al.},\ }\href@noop {} {\enquote {\bibinfo {title}
  {{Noble Element Simulation Technique v2.0}},}\ }\bibinfo {note}
  {\href{https://doi.org/10.5281/zenodo.1314669}{10.5281/zenodo.1314669
  (2018)}}\BibitemShut {NoStop}%
\bibitem [{\citenamefont {Ma}\ \emph {et~al.}(2020)\citenamefont {Ma} \emph
  {et~al.}}]{Ma2020_RnDepletion}%
  \BibitemOpen
  \bibfield  {author} {\bibinfo {author} {\bibfnamefont {W.}~\bibnamefont {Ma}}
  \emph {et~al.},\ }\href {\doibase 10.1088/1748-0221/15/12/P12038} {\bibfield
  {journal} {\bibinfo  {journal} {J. Instrum.}\ }\textbf {\bibinfo {volume}
  {15}},\ \bibinfo {pages} {P12038} (\bibinfo {year} {2020})}\BibitemShut
  {NoStop}%
\bibitem [{\citenamefont {Collon}\ \emph {et~al.}(2004)\citenamefont {Collon},
  \citenamefont {Kutschera},\ and\ \citenamefont {Lu}}]{Collon2004_forKr}%
  \BibitemOpen
  \bibfield  {author} {\bibinfo {author} {\bibfnamefont {P.}~\bibnamefont
  {Collon}}, \bibinfo {author} {\bibfnamefont {W.}~\bibnamefont {Kutschera}}, \
  and\ \bibinfo {author} {\bibfnamefont {Z.-T.}\ \bibnamefont {Lu}},\ }\href
  {\doibase 10.1146/annurev.nucl.53.041002.110622} {\bibfield  {journal}
  {\bibinfo  {journal} {Annu. Rev. Nucl. Part. Sci.}\ }\textbf {\bibinfo
  {volume} {54}},\ \bibinfo {pages} {39} (\bibinfo {year} {2004})}\BibitemShut
  {NoStop}%
\bibitem [{\citenamefont {Lu}\ \emph {et~al.}(2024)\citenamefont {Lu} \emph
  {et~al.}}]{Lu2024_P4Solarpp}%
  \BibitemOpen
  \bibfield  {author} {\bibinfo {author} {\bibfnamefont {X.}~\bibnamefont {Lu}}
  \emph {et~al.} (\bibinfo {collaboration} {PandaX Collaboration}),\ }\href
  {\doibase 10.1088/1674-1137/ad582a} {\bibfield  {journal} {\bibinfo
  {journal} {Chinese Physics C}\ }\textbf {\bibinfo {volume} {48}},\ \bibinfo
  {pages} {091001} (\bibinfo {year} {2024})}\BibitemShut {NoStop}%
\bibitem [{\citenamefont {Chen}\ \emph {et~al.}(2017)\citenamefont {Chen},
  \citenamefont {Chi}, \citenamefont {Liu},\ and\ \citenamefont
  {Wu}}]{Chen2017_AtomicEffectSolarNu}%
  \BibitemOpen
  \bibfield  {author} {\bibinfo {author} {\bibfnamefont {J.-W.}\ \bibnamefont
  {Chen}}, \bibinfo {author} {\bibfnamefont {H.-C.}\ \bibnamefont {Chi}},
  \bibinfo {author} {\bibfnamefont {C.-P.}\ \bibnamefont {Liu}}, \ and\
  \bibinfo {author} {\bibfnamefont {C.-P.}\ \bibnamefont {Wu}},\ }\href
  {\doibase https://doi.org/10.1016/j.physletb.2017.10.029} {\bibfield
  {journal} {\bibinfo  {journal} {Phys. Lett. B}\ }\textbf {\bibinfo {volume}
  {774}},\ \bibinfo {pages} {656} (\bibinfo {year} {2017})}\BibitemShut
  {NoStop}%
\bibitem [{\citenamefont {Si}\ \emph {et~al.}(2022)\citenamefont {Si} \emph
  {et~al.}}]{Si2022_Xe136DBD}%
  \BibitemOpen
  \bibfield  {author} {\bibinfo {author} {\bibfnamefont {L.}~\bibnamefont {Si}}
  \emph {et~al.} (\bibinfo {collaboration} {PandaX Collaboration}),\ }\href
  {\doibase 10.34133/2022/9798721} {\bibfield  {journal} {\bibinfo  {journal}
  {Research}\ }\textbf {\bibinfo {volume} {2022}},\ \bibinfo {pages} {9798721}
  (\bibinfo {year} {2022})}\BibitemShut {NoStop}%
\bibitem [{\citenamefont {Aprile}\ \emph {et~al.}(2022)\citenamefont {Aprile}
  \emph {et~al.}}]{Aprile2022_Xe124136}%
  \BibitemOpen
  \bibfield  {author} {\bibinfo {author} {\bibfnamefont {E.}~\bibnamefont
  {Aprile}} \emph {et~al.} (\bibinfo {collaboration} {XENON Collaboration}),\
  }\href {\doibase 10.1103/PhysRevC.106.024328} {\bibfield  {journal} {\bibinfo
   {journal} {Phys. Rev. C}\ }\textbf {\bibinfo {volume} {106}},\ \bibinfo
  {pages} {024328} (\bibinfo {year} {2022})}\BibitemShut {NoStop}%
\bibitem [{\citenamefont {Chen}\ \emph {et~al.}(2021)\citenamefont {Chen},
  \citenamefont {Cheng}, \citenamefont {Fu}, \citenamefont {Giuliani},
  \citenamefont {Liu}, \citenamefont {Lu}, \citenamefont {Ji}, \citenamefont
  {Qian}, \citenamefont {Qiao}, \citenamefont {Wang}, \citenamefont {Xia},
  \citenamefont {Xie}, \citenamefont {Yao},\ and\ \citenamefont
  {Zhang}}]{Chen2021_BambooMC}%
  \BibitemOpen
  \bibfield  {author} {\bibinfo {author} {\bibfnamefont {X.}~\bibnamefont
  {Chen}}, \bibinfo {author} {\bibfnamefont {C.}~\bibnamefont {Cheng}},
  \bibinfo {author} {\bibfnamefont {M.}~\bibnamefont {Fu}}, \bibinfo {author}
  {\bibfnamefont {F.}~\bibnamefont {Giuliani}}, \bibinfo {author}
  {\bibfnamefont {J.}~\bibnamefont {Liu}}, \bibinfo {author} {\bibfnamefont
  {X.}~\bibnamefont {Lu}}, \bibinfo {author} {\bibfnamefont {X.}~\bibnamefont
  {Ji}}, \bibinfo {author} {\bibfnamefont {Z.}~\bibnamefont {Qian}}, \bibinfo
  {author} {\bibfnamefont {H.}~\bibnamefont {Qiao}}, \bibinfo {author}
  {\bibfnamefont {Q.}~\bibnamefont {Wang}}, \bibinfo {author} {\bibfnamefont
  {J.}~\bibnamefont {Xia}}, \bibinfo {author} {\bibfnamefont {P.}~\bibnamefont
  {Xie}}, \bibinfo {author} {\bibfnamefont {Y.}~\bibnamefont {Yao}}, \ and\
  \bibinfo {author} {\bibfnamefont {H.}~\bibnamefont {Zhang}},\ }\href
  {\doibase 10.1088/1748-0221/16/09/T09004} {\bibfield  {journal} {\bibinfo
  {journal} {J. Instrum.}\ }\textbf {\bibinfo {volume} {16}},\ \bibinfo {pages}
  {T09004} (\bibinfo {year} {2021})}\BibitemShut {NoStop}%
\bibitem [{\citenamefont {Zeng}\ \emph {et~al.}(2024)\citenamefont {Zeng} \emph
  {et~al.}}]{Zeng2024_P4ERPhys}%
  \BibitemOpen
  \bibfield  {author} {\bibinfo {author} {\bibfnamefont {X.}~\bibnamefont
  {Zeng}} \emph {et~al.} (\bibinfo {collaboration} {PandaX Collaboration}),\
  }\href {https://arxiv.org/abs/2408.07641} {\enquote {\bibinfo {title}
  {{Exploring New Physics with PandaX-4T Low Energy Electronic Recoil Data}},}\
  } (\bibinfo {year} {2024}),\ \Eprint {http://arxiv.org/abs/2408.07641}
  {arXiv:2408.07641 [hep-ex]} \BibitemShut {NoStop}%
\bibitem [{\citenamefont {Huang}\ \emph
  {et~al.}(2022{\natexlab{a}})\citenamefont {Huang} \emph
  {et~al.}}]{Huang2022_NeutronBG}%
  \BibitemOpen
  \bibfield  {author} {\bibinfo {author} {\bibfnamefont {Z.}~\bibnamefont
  {Huang}} \emph {et~al.} (\bibinfo {collaboration} {PandaX Collaboration}),\
  }\href {\doibase 10.1088/1674-1137/ac8539} {\bibfield  {journal} {\bibinfo
  {journal} {Chinese Physics C}\ }\textbf {\bibinfo {volume} {46}},\ \bibinfo
  {pages} {115001} (\bibinfo {year} {2022}{\natexlab{a}})}\BibitemShut
  {NoStop}%
\bibitem [{\citenamefont {Ruppin}\ \emph {et~al.}(2014)\citenamefont {Ruppin},
  \citenamefont {Billard}, \citenamefont {Figueroa-Feliciano},\ and\
  \citenamefont {Strigari}}]{Ruppin2014_NeutrinoBg}%
  \BibitemOpen
  \bibfield  {author} {\bibinfo {author} {\bibfnamefont {F.}~\bibnamefont
  {Ruppin}}, \bibinfo {author} {\bibfnamefont {J.}~\bibnamefont {Billard}},
  \bibinfo {author} {\bibfnamefont {E.}~\bibnamefont {Figueroa-Feliciano}}, \
  and\ \bibinfo {author} {\bibfnamefont {L.}~\bibnamefont {Strigari}},\ }\href
  {\doibase 10.1103/PhysRevD.90.083510} {\bibfield  {journal} {\bibinfo
  {journal} {Phys. Rev. D}\ }\textbf {\bibinfo {volume} {90}},\ \bibinfo
  {pages} {083510} (\bibinfo {year} {2014})}\BibitemShut {NoStop}%
\bibitem [{\citenamefont {Abdukerim}\ \emph {et~al.}(2022)\citenamefont
  {Abdukerim} \emph {et~al.}}]{Abdukerim2022_AC}%
  \BibitemOpen
  \bibfield  {author} {\bibinfo {author} {\bibfnamefont {A.}~\bibnamefont
  {Abdukerim}} \emph {et~al.},\ }\href {\doibase 10.1088/1674-1137/ac7cd8}
  {\bibfield  {journal} {\bibinfo  {journal} {Chinese Physics C}\ }\textbf
  {\bibinfo {volume} {46}},\ \bibinfo {pages} {103001} (\bibinfo {year}
  {2022})}\BibitemShut {NoStop}%
\bibitem [{Note1()}]{Note1}%
  \BibitemOpen
  \bibinfo {note} {These values are updated after unblinding, see
  later.}\BibitemShut {Stop}%
\bibitem [{\citenamefont {Cowan}()}]{Cowan2012_DiscoverySF}%
  \BibitemOpen
  \bibfield  {author} {\bibinfo {author} {\bibfnamefont {G.}~\bibnamefont
  {Cowan}},\ }\href@noop {} {\enquote {\bibinfo {title} {{Discovery sensitivity
  for a counting experiment with background uncertainty}},}\ }\bibinfo {note}
  {\href{https://www.pp.rhul.ac.uk/~cowan/stat/medsig/medsigNote.pdf}{www.pp.rhul.ac.uk/~cowan/stat/medsig/medsigNote.pdf}}\BibitemShut
  {NoStop}%
\bibitem [{\citenamefont {Baxter}\ \emph {et~al.}(2021)\citenamefont {Baxter}
  \emph {et~al.}}]{Baxter2021_ReportResWhitePaper}%
  \BibitemOpen
  \bibfield  {author} {\bibinfo {author} {\bibfnamefont {D.}~\bibnamefont
  {Baxter}} \emph {et~al.},\ }\href@noop {} {\bibfield  {journal} {\bibinfo
  {journal} {Eur. Phys. J. C}\ }\textbf {\bibinfo {volume} {81}} (\bibinfo
  {year} {2021})}\BibitemShut {NoStop}%
\bibitem [{\citenamefont {Cowan}\ \emph
  {et~al.}(2011{\natexlab{a}})\citenamefont {Cowan}, \citenamefont {Cranmer},
  \citenamefont {Gross},\ and\ \citenamefont {Vitells}}]{Cowan:2010js}%
  \BibitemOpen
  \bibfield  {author} {\bibinfo {author} {\bibfnamefont {G.}~\bibnamefont
  {Cowan}}, \bibinfo {author} {\bibfnamefont {K.}~\bibnamefont {Cranmer}},
  \bibinfo {author} {\bibfnamefont {E.}~\bibnamefont {Gross}}, \ and\ \bibinfo
  {author} {\bibfnamefont {O.}~\bibnamefont {Vitells}},\ }\href {\doibase
  10.1140/epjc/s10052-011-1554-0} {\bibfield  {journal} {\bibinfo  {journal}
  {Eur. Phys. J. C}\ }\textbf {\bibinfo {volume} {71}},\ \bibinfo {pages}
  {1554} (\bibinfo {year} {2011}{\natexlab{a}})},\ \bibinfo {note} {[Erratum:
  Eur.Phys.J.C 73, 2501 (2013)]}\BibitemShut {NoStop}%
\bibitem [{\citenamefont {Gross}\ and\ \citenamefont
  {Vitells}(2010)}]{Gross2010_LEE}%
  \BibitemOpen
  \bibfield  {author} {\bibinfo {author} {\bibfnamefont {E.}~\bibnamefont
  {Gross}}\ and\ \bibinfo {author} {\bibfnamefont {O.}~\bibnamefont
  {Vitells}},\ }\href {\doibase 10.1140/epjc/s10052-010-1470-8} {\bibfield
  {journal} {\bibinfo  {journal} {Eur. Phys. J. C}\ }\textbf {\bibinfo {volume}
  {70}},\ \bibinfo {pages} {525} (\bibinfo {year} {2010})}\BibitemShut
  {NoStop}%
\bibitem [{\citenamefont {Cowan}\ \emph
  {et~al.}(2011{\natexlab{b}})\citenamefont {Cowan}, \citenamefont {Cranmer},
  \citenamefont {Gross},\ and\ \citenamefont {Vitells}}]{Cowan:2011an}%
  \BibitemOpen
  \bibfield  {author} {\bibinfo {author} {\bibfnamefont {G.}~\bibnamefont
  {Cowan}}, \bibinfo {author} {\bibfnamefont {K.}~\bibnamefont {Cranmer}},
  \bibinfo {author} {\bibfnamefont {E.}~\bibnamefont {Gross}}, \ and\ \bibinfo
  {author} {\bibfnamefont {O.}~\bibnamefont {Vitells}},\ }\href@noop {} {\
  (\bibinfo {year} {2011}{\natexlab{b}})},\ \Eprint
  {http://arxiv.org/abs/1105.3166} {arXiv:1105.3166 [physics.data-an]}
  \BibitemShut {NoStop}%
\bibitem [{Note2()}]{Note2}%
  \BibitemOpen
  \bibinfo {note} {See Supplemental Material at \protect \url
  {http://link.aps.org/supplemental/10.1103/PhysRevLett.134.011805}, which
  includes Refs.~\cite {Klos2013_ChiralEFT, Amole2019_PICOSDp,
  atlas2018_2HDMa_propo, Huang2022_P4SD, atlas2023_2HDMa, Li2018_DMSimplified,
  Li2019_DMSimplifiedLoopEffects}, for results on alternative models and more
  information on the statistic test}\BibitemShut {NoStop}%
\bibitem [{\citenamefont {Klos}\ \emph {et~al.}(2013)\citenamefont {Klos},
  \citenamefont {Men\'endez}, \citenamefont {Gazit},\ and\ \citenamefont
  {Schwenk}}]{Klos2013_ChiralEFT}%
  \BibitemOpen
  \bibfield  {author} {\bibinfo {author} {\bibfnamefont {P.}~\bibnamefont
  {Klos}}, \bibinfo {author} {\bibfnamefont {J.}~\bibnamefont {Men\'endez}},
  \bibinfo {author} {\bibfnamefont {D.}~\bibnamefont {Gazit}}, \ and\ \bibinfo
  {author} {\bibfnamefont {A.}~\bibnamefont {Schwenk}},\ }\href {\doibase
  10.1103/PhysRevD.88.083516} {\bibfield  {journal} {\bibinfo  {journal} {Phys.
  Rev. D}\ }\textbf {\bibinfo {volume} {88}},\ \bibinfo {pages} {083516}
  (\bibinfo {year} {2013})}\BibitemShut {NoStop}%
\bibitem [{\citenamefont {Amole}\ \emph {et~al.}(2019)\citenamefont {Amole}
  \emph {et~al.}}]{Amole2019_PICOSDp}%
  \BibitemOpen
  \bibfield  {author} {\bibinfo {author} {\bibfnamefont {C.}~\bibnamefont
  {Amole}} \emph {et~al.} (\bibinfo {collaboration} {PICO Collaboration}),\
  }\href {\doibase 10.1103/PhysRevD.100.022001} {\bibfield  {journal} {\bibinfo
   {journal} {Phys. Rev. D}\ }\textbf {\bibinfo {volume} {100}},\ \bibinfo
  {pages} {022001} (\bibinfo {year} {2019})}\BibitemShut {NoStop}%
\bibitem [{\citenamefont {Aad}\ \emph {et~al.}(2024)\citenamefont {Aad} \emph
  {et~al.}}]{atlas2023_2HDMa}%
  \BibitemOpen
  \bibfield  {author} {\bibinfo {author} {\bibfnamefont {G.}~\bibnamefont
  {Aad}} \emph {et~al.} (\bibinfo {collaboration} {ATLAS Collaboration}),\
  }\href {\doibase https://doi.org/10.1016/j.scib.2024.06.003} {\bibfield
  {journal} {\bibinfo  {journal} {Sci. Bull.}\ }\textbf {\bibinfo {volume}
  {69}},\ \bibinfo {pages} {3005} (\bibinfo {year} {2024})}\BibitemShut
  {NoStop}%
\bibitem [{\citenamefont {Abe}\ \emph {et~al.}(2020)\citenamefont {Abe} \emph
  {et~al.}}]{atlas2018_2HDMa_propo}%
  \BibitemOpen
  \bibfield  {author} {\bibinfo {author} {\bibfnamefont {T.}~\bibnamefont
  {Abe}} \emph {et~al.} (\bibinfo {collaboration} {ATLAS Collaboration}),\
  }\href {\doibase https://doi.org/10.1016/j.dark.2019.100351} {\bibfield
  {journal} {\bibinfo  {journal} {Phys. Dark Univ.}\ }\textbf {\bibinfo
  {volume} {27}},\ \bibinfo {pages} {100351} (\bibinfo {year}
  {2020})}\BibitemShut {NoStop}%
\bibitem [{\citenamefont {Huang}\ \emph
  {et~al.}(2022{\natexlab{b}})\citenamefont {Huang} \emph
  {et~al.}}]{Huang2022_P4SD}%
  \BibitemOpen
  \bibfield  {author} {\bibinfo {author} {\bibfnamefont {Z.}~\bibnamefont
  {Huang}} \emph {et~al.} (\bibinfo {collaboration} {PandaX Collaboration}),\
  }\href {\doibase 10.1016/j.physletb.2022.137487} {\bibfield  {journal}
  {\bibinfo  {journal} {Phys. Lett. B}\ }\textbf {\bibinfo {volume} {834}},\
  \bibinfo {pages} {137487} (\bibinfo {year} {2022}{\natexlab{b}})}\BibitemShut
  {NoStop}%
\bibitem [{\citenamefont {Li}(2018)}]{Li2018_DMSimplified}%
  \BibitemOpen
  \bibfield  {author} {\bibinfo {author} {\bibfnamefont {T.}~\bibnamefont
  {Li}},\ }\href {\doibase 10.1016/j.physletb.2018.05.073} {\bibfield
  {journal} {\bibinfo  {journal} {Phys. Lett. B}\ }\textbf {\bibinfo {volume}
  {782}},\ \bibinfo {pages} {497} (\bibinfo {year} {2018})}\BibitemShut
  {NoStop}%
\bibitem [{\citenamefont {Li}\ and\ \citenamefont
  {Wu}(2019)}]{Li2019_DMSimplifiedLoopEffects}%
  \BibitemOpen
  \bibfield  {author} {\bibinfo {author} {\bibfnamefont {T.}~\bibnamefont
  {Li}}\ and\ \bibinfo {author} {\bibfnamefont {P.}~\bibnamefont {Wu}},\ }\href
  {\doibase 10.1088/1674-1137/43/11/113102} {\bibfield  {journal} {\bibinfo
  {journal} {Chin. Phys. C}\ }\textbf {\bibinfo {volume} {43}},\ \bibinfo
  {pages} {113102} (\bibinfo {year} {2019})}\BibitemShut {NoStop}%
\end{thebibliography}%

\end{document}